\documentclass[12pt]{article}
\usepackage{amsmath}
\usepackage{txfonts,bm,pifont}  
\usepackage{cite}  
\usepackage{graphicx,xcolor}
\usepackage[dvipdfm,bookmarks=true,bookmarksnumbered=true,colorlinks=true]{hyperref}
\newtheorem{theorem}{Theorem}[section]
\newtheorem{proposition}[theorem]{Proposition}
\newtheorem{corollary}[theorem]{Corollary}
\newtheorem{lemma}[theorem]{Lemma}
\newtheorem{remark}[theorem]{Remark}
\newtheorem{definition}[theorem]{Definition}

\newcommand{\taus}{\tau^*{}}
\makeatletter
\@addtoreset{equation}{section}
\renewcommand{\theequation}{\@arabic\c@section.\@arabic\c@equation}
\long\def\@makecaption#1#2{
 \vskip 10pt 
 \setbox\@tempboxa\hbox{#1. #2}
 \ifdim \wd\@tempboxa >\hsize #1. #2\par \else \hbox
to\hsize{\hfil\box\@tempboxa\hfil} 
 \fi}
\makeatother
\begin{document}
\begin{center}
\renewcommand{\baselinestretch}{1.3}\selectfont
\begin{large}
\textbf{MOTION AND B\"ACKLUND TRANSFORMATIONS OF DISCRETE PLANE CURVES}
\end{large}\\[4mm]
\renewcommand{\baselinestretch}{1}\selectfont
\textrm{\large Jun-ichi Inoguchi$^1$, Kenji Kajiwara$^2$,
Nozomu Matsuura$^3$ and Yasuhiro Ohta$^4$ }\\[2mm]
$^1$: Department of Mathematical Sciences, Yamagata University,\\
 1-4-12 Kojirakawa-machi, Yamagata 990-8560, Japan.\\
 inoguchi@sci.kj.yamagata-u.ac.jp\\
$^2$: Institute of Mathematics for Industry, Kyushu University, 744 Motooka, Fukuoka 819-8581, Japan.\\
kaji@imi.kyushu-u.ac.jp
\\
$^3$: Department of Applied Mathematics, Fukuoka University,\\
 Nanakuma, Fukuoka 814-0180, Japan.\\
nozomu@fukuoka-u.ac.jp 
\\
 $^4$: Department of Mathematics, Kobe University, Rokko, Kobe 657-8501, Japan.
\\
ohta@math.sci.kobe-u.ac.jp
\\[2mm]
3 March 2011; Revised on 4 October 2011
\end{center}
\begin{abstract}
We construct explicit solutions to the discrete motion of discrete plane
curves that has been introduced by one of the authors
recently. Explicit formulas in terms the $\tau$ function are presented.
Transformation theory of the motions of both smooth and discrete curves is developed simultaneously.
\end{abstract}



\section{Introduction}

Differential geometry has a close relationship with the theory of integrable systems. In fact, many
integrable differential or difference equations arise as compatibility conditions of some
geometric objects. For instance, it is well known that the compatibility condition of
pseudospherical surfaces gives rise to the sine-Gordon equation under the Chebyshev net
parametrization. For more information on such connections we refer to a monograph
\cite{Rogers-Schief} by Rogers and Schief.

The above connection between the differential geometry of surfaces and integrable systems has
been known since the nineteenth century (although the theory of integrable systems was not yet
established). However, it is curious that the link between the differential geometry of curves and
integrable systems has been noticed rather recently. Actually Lamb\cite{Lamb} and
Goldstein and Petrich \cite{Goldstein-Petrich} discovered an interesting connection between integrable
systems and the differential geometry of plane curves. Namely, they found that the \textit{modified
Korteweg--de Vries equation} (mKdV equation) appears as the compatibility condition of
a certain motion of plane curves. Here a motion of curves means an isoperimetric time evolution of
arc-length parametrized plane curves.  More precisely, the compatibility condition implies that the
curvature function of a motion should satisfy the mKdV equation. As a result, the angle function of
the motion satisfies the \textit{potential modified Korteweg--de Vries equation} (potential mKdV
equation).

On the other hand, in the theory of integrable systems, much attention has been paid to
discretization of integrable differential equations preserving integrability, after the
pioneering work of Ablowitz and Ladik\cite{Ablowitz:book} and
Hirota\cite{Hirota:difference1,Hirota:difference2,Hirota:dsG,Hirota:difference4,Hirota:difference5}.
Later, Date, Jimbo and Miwa developed a unified algebraic approach from the view of so-called the KP
theory\cite{DJM:discrete1,DJM:discrete2,DJM:discrete3,DJM:discrete4,DJM:discrete5,Jimbo-Miwa,Miwa}.
For other approaches to discrete integrable systems, see, for example, \cite{Suris:book,
Nijhoff}.  Thus one can expect the existence of discretized differential geometric objects governed
by discrete integrable systems. This idea has been realized by the works of
Bobenko and Pinkall\cite{BP} and Doliwa\cite{Doliwa:dToda} where the discrete analogue of classical
surface theory has been proposed, and it is now actively studied under the name of discrete
differential geometry\cite{Bobenko-Suris}.

In the case of the discrete analogue of curves, Doliwa and Santini studied continuous motion of discrete
curves in the 3-sphere, and obtained the Ablowitz--Ladik hierarchy in \cite{DS1}, where a semi-discrete
(discrete in space variable and continuous in time variable) analogue of the mKdV equation is derived as
the simplest case. Their formulation includes the plane curves as a limiting case. Hisakado {\it et al}
proposed a discretization of arc-length parametrized plane curves\cite{HNW}, and obtained another
semi-discretization of the mKdV equation. Hoffmann and Kutz\cite{HK,Hoffmann:LN} considered
discretization of the curvature function. By using their discrete curvature function and M\"obius
geometry, they obtained a semi-discrete mKdV equation that is the same as the one in \cite{DS1}.

The discretization of the time variable of discrete curve motion in the 3-sphere was studied in
\cite{DS2,DS3}, where the evolution corresponded to the discrete sine-Gordon
equation\cite{Hirota:dsG}. As is well known, binormal motion of space curves induces the nonlinear
Schr\"odinger equation via Hasimoto transformation \cite{Hasimoto,Lamb}. Discretization of this
curve motion has been formulated in \cite{Hoffmann:dNLS,Pinkall:dNLS}.

Recently one of the authors of the present paper formulated a full discretization of the motion of
plane discrete curves\cite{Matsuura} in a purely Euclidean geometric manner, where the discrete
potential mKdV equation proposed by Hirota\cite{Hirota:dpmKdV} is deduced as the compatibility
condition.  It admits a natural continuous limit to the potential mKdV equation describing
continuous motion of a smooth plane curves.

In the smooth curve theory, the potential function coincides with the angle function of a curve, a
primitive function of the curvature. However, in the discrete case, the potential function and the
angle function become different objects. In this framework, the primal geometric object is the
potential function rather than the curvature (see \cite{DS2,Matsuura} and Section 2 of the present
paper). Natural and systematic construction of the discrete motion of the curves is expected by
using the theory of discrete integrable systems.

The purpose of this paper is to construct explicit solutions to discrete motion of discrete plane
curves by using the theory of $\tau$ functions. This paper is organized as follows.  In Section 2, we
prepare fundamental ingredients of plane curve geometry and motions (isoperimetric time evolutions)
of plane curves described by the potential mKdV equation. Next we give a brief review of the
discrete motion of discrete curves\cite{Matsuura}. In Section 3, we shall give a construction of
motions for both smooth and discrete curves by the theory of $\tau$ functions.  More precisely we
introduce a system of bilinear equations of Hirota type, which can be obtained by a certain reduction
of the discrete two-dimensional Toda lattice hierarchy\cite{Jimbo-Miwa,Tsujimoto,Ueno-Takasaki}. We
shall give a representation formula for curve motions in terms of the $\tau$ function.

One of the central topics in classical differential geometry is the transformation theory of curves
and surfaces. The best known example might be the B\"acklund transformations of pseudospherical
surfaces. The original B\"acklund transformation was defined as a tangential line congruence
satisfying the \textit{constant distance property} and \textit{constant normal angle property} (see
\cite{Rogers-Schief} ). In plane curve geometry, B\"acklund transformations on arc-length
parametrized plane curves can be defined as arc-length preserving transformations satisfying
the constant distance property. Such transformations can be extended to transformations on smooth curve
motions via the transformation of solutions to the potential mKdV equation. Motivated by this fact,
we shall introduce B\"acklund transformations for discrete motion of discrete curves in Section 4
(compare with \cite{Hoffmann:LN}). In particular we shall give another type of B\"acklund
transformation on motions of both smooth and discrete curves, which is related to the discrete
sine-Gordon equation. In Section 5, we shall construct and exhibit some explicit solutions of curve
motions, namely, the multi-soliton and multi-breather solutions. We also present some pictures of
discrete motions of discrete curves. We finally give some explicit formulas for the B\"acklund
transformations of both smooth and discrete curve motions via the $\tau$ functions.

\section{Motion of plane curves}
Let $\gamma(x)$ be an arc-length parametrized curve in the Euclidean plane
$\mathbb{R}^2$. Then the Frenet equation of $\gamma$ is 
\begin{equation}
 \gamma''=\left[\begin{array}{cc} 0&-\kappa \\\kappa & 0\end{array}\right]\gamma'.
\label{gamma_x}
\end{equation}
Here $'$ denotes the differentiation with respect to $x$, and the function $\kappa$ is the
\textit{curvature} of $\gamma$. Let us consider the following motion in time $t$, i.e.,
isoperimetric time evolution:
\begin{equation}
 \frac{\partial}{\partial t}\gamma'
=\left[
\begin{array}{cc}0 &\kappa''+\dfrac{\kappa^3}{2} \\-\kappa''-\dfrac{\kappa^3}{2} &0\end{array}
\right]\gamma'.
\label{gamma_t}
\end{equation}
Then the \textit{potential function} $\theta(x,t)$ defined by $\kappa=\theta'$ satisfies the
\textit{potential mKdV equation}\cite{Goldstein-Petrich,Lamb}:
\begin{equation}
 \theta_t + \frac{1}{2}(\theta_x)^3 + \theta_{xxx}=0.\label{eqn:pmKdV}
\end{equation}
The function $\theta$ is called the \textit{angle function} of $\gamma$
in differential geometry. Note that $\gamma'$ can be expressed as
\begin{equation}
 \gamma'=\left[\begin{array}{l}\cos\theta  \\\sin\theta\end{array}\right].\label{angle_function}
\end{equation}
For any non-zero constant $\lambda$, the set of equations
\begin{align}
 &\frac{\partial}{\partial x}\left(\frac{\widetilde\theta +\theta}{2}\right)
=2\lambda\sin\frac{\widetilde\theta-\theta}{2},\label{BT1:pmKdV}\\[1mm]
 &\frac{\partial}{\partial t}\left(\frac{\widetilde\theta +\theta}{2}\right)
=-\lambda\left\{\left(\theta_x\right)^2+8\lambda^2\right\}
\sin\frac{\widetilde\theta-\theta}{2}+2\lambda\theta_{xx}\cos\frac{\widetilde\theta-\theta}{2}
+4\lambda^2\theta_x,\label{BT2:pmKdV}
\end{align}
defines  a solution $\widetilde\theta$ to the potential mKdV equation\cite{Wadati}. The
solution $\widetilde\theta$ is called a \textit{B\"acklund transform} of $\theta$.

\begin{definition}\rm
 A map $\gamma:~\mathbb{Z}\rightarrow \mathbb{R}^2;~n\mapsto \gamma_n$ is said to be a
 \textit{discrete curve} of segment length $a_n$ if
\begin{equation}
 \left|\frac{\gamma_{n+1} - \gamma_n}{a_n}\right| = 1.
\label{iso0}
\end{equation}
\end{definition}
\par\bigskip
\noindent We introduce the \textit{angle function} $\Psi_n$ of a discrete curve $\gamma$ by
\begin{equation}
 \frac{\gamma_{n+1} - \gamma_n}{a_n}=\left[\begin{array}{l}\cos\Psi_n \\\sin\Psi_n \end{array}\right].\label{discrete_angle}
\end{equation}
A discrete curve $\gamma$ satisfies
\begin{equation}
 \frac{\gamma_{n+1} - \gamma_{n}}{a_n} = R(K_n)~\frac{\gamma_{n} - \gamma_{n-1}}{a_{n-1}},
\end{equation}
for $K_n=\Psi_{n}-\Psi_{n-1}$, where $R(K_n)$ denotes the rotation matrix given by
\begin{equation}
 R(K_n)=\left(\begin{array}{cc}\cos K_n & -\sin K_n\\\sin K_n & \cos K_n\end{array}\right).
\end{equation}

Now let us recall the following discrete motion of discrete curve
$\gamma_n^m:\ \mathbb{Z}^{2}\rightarrow \mathbb{R}^2$ introduced by Matsuura\cite{Matsuura}:
\begin{align}
& \left|\frac{\gamma^{m}_{n+1} - \gamma^{m}_n}{a_n}\right| = 1,
\label{iso}\\[2mm]
& \frac{\gamma^{m}_{n+1} - \gamma^{m}_{n}}{a_n} = R(K^m_n)~\frac{\gamma^{m}_{n} - \gamma^{m}_{n-1}}{a_{n-1}},\label{gamma_n}\\[2mm]
& \frac{\gamma^{m+1}_{n} - \gamma^{m}_{n}}{b_m} = R(W^m_n)~\frac{\gamma^{m}_{n+1} - \gamma^{m}_{n}}{a_{n}},
\label{gamma_m}
\end{align}
where $a_n$ and $b_m$ are arbitrary functions of $n$ and $m$,
respectively. Compatibility of the system (\ref{iso})--(\ref{gamma_m}) implies the existence of
the \textit{potential function} $\Theta_n^m$ defined by
\begin{equation}
W^m_n =\frac{\Theta^{m+1}_{n} - \Theta^{m}_{n+1}}{2}, \quad
K^m_n = \frac{\Theta^{m}_{n+1} - \Theta^{m}_{n-1}}{2},
\end{equation}
and it follows that $\Theta_n^m$ satisfies the \textit{discrete potential mKdV equation}\cite{Hirota:dpmKdV}:
\begin{equation}
  \tan\left(\frac{\Theta_{n+1}^{m+1}-\Theta_{n}^m}{4}\right) = \frac{b_m+a_n}{b_m-a_n}~\tan\left(\frac{\Theta_{n}^{m+1} - \Theta_{n+1}^{m}}{4}\right).\label{eqn:dpmKdV} 
\end{equation}
Note that the angle function $\Psi_n^m$ can be expressed as
\begin{equation}
 \Psi_n^m = \frac{\Theta_{n+1}^m+\Theta_n^m}{2}.\label{discrete_angle_and_potential}
\end{equation}
\begin{remark}\rm
The potential discrete mKdV equation (\ref{eqn:dpmKdV}) has also been known as the superposition
formula for the modified KdV equation (\ref{eqn:pmKdV})\cite{Wadati} and the sine-Gordon equation \cite{Bianchi,Rogers-Schief}.
\end{remark}
\section{The $\tau$ function representation of plane curves}
In this section, we give a representation formula for curve motions in terms of $\tau$ functions.

Let $\tau_n^m=\tau_n^m(x,t;y)$ be a complex-valued function dependent on two discrete variables $m$
and $n$, and three continuous variables $x$, $t$ and $y$, which satisfies the following system of bilinear equations:
\begin{align}
 & \frac{1}{2}D_xD_y~\tau_n^m\cdot\tau_n^m = -\left(\tau^*{}_n^m\right)^2,\label{bl1}\\
 & D_x^2~\tau_n^m\cdot\taus_n^m =0,\label{bl2}\\
 &\left(D_x^3+D_t\right)~\tau_n^m\cdot\taus_n^m = 0,\label{bl3}\\
 & D_y~\tau_{n+1}^m\cdot\tau_n^m = -a_n\tau^*{}_{n+1}^m\tau^*{}_n^m,\label{bl4}\\
 & D_y~\tau_{n}^{m+1}\cdot\tau_n^m = -b_m\tau^*{}_{n+1}^m\tau^*{}_n^m,\label{bl5}\\
 &b_m\tau^*{}_{n}^{m+1}\tau_{n+1}^m - a_n\tau^*{}_{n+1}^{m}\tau_{n}^{m+1}
+ (a_n-b_m)\tau^*{}_{n+1}^{m+1}\tau_n^m=0.\label{bl6}
\end{align}
Here, ${}^*$ denotes the complex conjugate, and $D_x$, $D_y$ and $D_t$ are
Hirota's \textit{bilinear differential operators} ($D$-operators) defined by
\begin{equation}
 D_x^iD_y^jD_t^k~f\cdot g = \left.(\partial_x-\partial_{x'})^i(\partial_y-\partial_{y'})^j(\partial_t-\partial_{t'})^k~f(x,y,t)g(x',y',t')\right|_{x=x',y=y',t=t'}~.
\end{equation}
For the calculus of the $D$-operators, we refer to \cite{Hirota:book}.  In general, the functions
satisfying the bilinear equations of Hirota type are called \textit{$\tau$ functions}.
\begin{theorem}\label{thm:tau_formula}
 Let $\tau_n^m$ be a solution to equations (\ref{bl1})--(\ref{bl6}). Define a real function
$\Theta_n^m(x,t;y)$ and an $\mathbb{R}^2$-valued function $\gamma_n^m(x,t;y)$ by
\begin{align}
& \Theta_n^m(x,t;y):=\frac{2}{\sqrt{-1}}\log\frac{\tau_n^m}{\tau^*{}_n^m},\\
&\gamma_n^m(x,t;y):=\left[\begin{array}{c} 
{\displaystyle -\frac{1}{2}\left(\log \tau_n^m\tau^*{}_n^m\right)_y}\\[2mm]
{\displaystyle \frac{1}{2\sqrt{-1}}\left(\log \frac{\tau_n^m}{\tau^*{}_n^m}\right)_y}
\end{array}\right].
\end{align}
\begin{enumerate}
 \item For any $m,n\in\mathbb{Z}$ and $y\in\mathbb{R}$, the functions
       $\theta(x,t)=\Theta_n^m(x,t;y)$ and
$\gamma(x,t)=\gamma_n^m(x,t;y)$ satisfy equations (\ref{gamma_x})--(\ref{eqn:pmKdV}).
 \item For any $x,t,y\in\mathbb{R}$, the functions $\Theta_n^m=\Theta_n^m(x,t;y)$ and
$\gamma_n^m=\gamma_n^m(x,t;y)$ satisfy equations (\ref{iso})--(\ref{eqn:dpmKdV}).
\end{enumerate}
\end{theorem}
\noindent\textbf{Proof.}\quad (1) Express $\gamma_n^m={}^t(X_n^m,Y_n^m)$. Then by using equation (\ref{bl1})
together with its complex conjugate, we have
\begin{eqnarray*}
\left(X^{m}_n\right)^\prime &=& - \frac{1}{2}\log(\taus^m_n\tau^m_n)_{xy}
= - \frac{1}{2}\left[\frac{\frac{1}{2}D_xD_y~\taus^m_n\cdot\taus^m_n}{(\taus^m_n)^2} 
+ \frac{\frac{1}{2}D_xD_y~\tau^m_n\cdot\tau^m_n}{(\tau^m_n)^2}\right]\\
&=&  \frac{1}{2}\left[\left(\frac{\tau_n^m}{\taus^m_n}\right)^2 
+\left(\frac{\taus_n^m}{\tau^m_n}\right)^2\right]=
\cos\Theta_n^m.
\end{eqnarray*}
Similarly we obtain $ \left(Y^{m}_n\right)^\prime=\sin\Theta_n^m$.
Differentiating $(\gamma_n^m)'={}^{{\rm t}}(\cos\Theta_n^m,\sin\Theta_n^m)$ by
$x$ and noticing that $\kappa=\Theta'$ , we obtain equation (\ref{gamma_x}):

\begin{displaymath}
 (\gamma_n^m)'' = \left(\Theta_n^m\right)'\left(\begin{array}{c}-\sin\Theta_n^m \\ \cos\Theta_n^m\end{array}\right)
= \left(\begin{array}{cc}0  & -\kappa\\ \kappa & 0\end{array}\right)(\gamma_n^m)'.
\end{displaymath}
On the other hand, differentiating $(\gamma_n^m)'$ by $t$, we have
\begin{displaymath}
 (\gamma_n^m)'_t = \left(\Theta_n^m\right)_t\left(\begin{array}{c}-\sin\Theta_n^m \\ \cos\Theta_n^m\end{array}\right)
= \left(\Theta_n^m\right)_t\left(\begin{array}{cc}0  & -1\\ 1 & 0\end{array}\right)(\gamma_n^m)'.
\end{displaymath}
By using the bilinear equations (\ref{bl2}) and (\ref{bl3}), $(\Theta_n^m)_t$ can be rewritten as
\begin{eqnarray}
(\Theta_n^m)_t&=&\frac{2}{\sqrt{-1}}\frac{D_t~\tau_n^m\cdot\taus_n^m}{\tau_n^m\taus_n^m}
=-\frac{2}{\sqrt{-1}}\frac{D_x^3~\tau_n^m\cdot\taus_n^m}{\tau_n^m\taus_n^m} \nonumber\\
&=&-\frac{2}{\sqrt{-1}}\left[\left(\log\frac{\tau_n^m}{\taus_n^m}\right)_{xxx} 
+ 3\left(\log\frac{\tau_n^m}{\taus_n^m}\right)_{x}
\left(\log\tau_n^m\taus_n^m\right)_{xx}
+ \left\{\left(\log\frac{\tau_n^m}{\taus_n^m}\right)_{x} \right\}^3\right]\nonumber\\[2mm]
&=&-\frac{2}{\sqrt{-1}}\left[\left(\log\frac{\tau_n^m}{\taus_n^m}\right)_{xxx} 
-2 \left\{\left(\log\frac{\tau_n^m}{\taus_n^m}\right)_{x} \right\}^3\right]
=-\kappa_{xx} - \frac{\kappa^3}{2}, \label{th1:proof_kappa}
\end{eqnarray}
which yields equation (\ref{gamma_t}). Here we have used the relation
\begin{displaymath}
0 = \frac{D_x^2~\tau_n^m\cdot\taus_n^m}{\tau_n^m\taus_n^m}
= \left(\log\tau_n^m\taus_n^m\right)_{xx} + \left(\log\frac{\tau_n^m}{\taus_n^m}\right)_{x}^2,
\end{displaymath}
which is a consequence of equation (\ref{bl2}).
The potential mKdV equation (\ref{eqn:pmKdV}) follows immediately from equation (\ref{th1:proof_kappa})
by noticing that $\kappa=\Theta'$. 

(2) From equation (\ref{bl4}) and its complex conjugate we have
\begin{equation}
\left(\log\frac{\tau_{n+1}^m}{\tau_n^m}\right)_y = -a_n~\frac{\taus_{n+1}^m\taus_n^m}{\tau_{n+1}^m\tau_n^m},\qquad
\left(\log\frac{\taus_{n+1}^m}{\taus_n^m}\right)_y  = -a_n~\frac{\tau_{n+1}^m\tau_n^m}{\taus_{n+1}^m\taus_n^m}.\label{th1:proof1}
\end{equation}
Adding these two equations we obtain
\begin{equation}
 \left(\log\tau_{n+1}^m\taus_{n+1}^m\right)_y -  \left(\log\tau_{n}^m\taus_{n}^m\right)_y 
=-a_n\left(\frac{\taus_{n+1}^m\taus_n^m}{\tau_{n+1}^m\tau_n^m}
+ \frac{\tau_{n+1}^m\tau_n^m}{\taus_{n+1}^m\taus_n^m}
\right),
\end{equation}
which yields
\begin{equation}
\frac{X_{n+1}^m - X_n^m}{a_n} = 
\cos\Psi_n^m,\quad \Psi_n^m 
= \frac{1}{\sqrt{-1}}
\log\left(\frac{\tau_{n+1}^m\tau_n^m}{\taus_{n+1}^m\taus_n^m}\right)
= \frac{\Theta_{n+1}^m  + \Theta_{n}^m }{2}.
\end{equation}
Subtracting the second equation from the first equation in
(\ref{th1:proof1}) we have
\begin{eqnarray*}
\frac{Y_{n+1}^m{} - Y_n^m}{a_n} = \sin\Psi_n^m.
\end{eqnarray*}
Therefore we obtain
\begin{equation}
\frac{\gamma_{n+1}^m -\gamma_n^m}{a_n}
= \left(\begin{array}{c}\smallskip\cos\Psi_n^m \\\sin\Psi_n^m \end{array}\right),\label{segment_n}
\end{equation}
which gives equation (\ref{iso}). Next, from equation (\ref{segment_n}) we see that 
\begin{equation}
\frac{\gamma_{n+1}^m -\gamma_n^m}{a_n}
=R(\Psi_n^m - \Psi_{n-1}^m)~\frac{\gamma_{n}^m -\gamma_{n-1}^m}{a_{n-1}},\quad
\Psi_n^m - \Psi_{n-1}^m =\frac{\Theta_{n+1}^m-\Theta_{n-1}^m}{2}= K_n^m,
\end{equation}
which is nothing but equation (\ref{gamma_n}). 
Similarly, starting from equation (\ref{bl5}) and its complex conjugate we obtain
\begin{equation}
\frac{\gamma_{n}^{m+1} -\gamma_n^m}{b_m}= \left(\begin{array}{c}\smallskip\cos\Phi_n^m \\\sin\Phi_n^m \end{array}\right),\qquad
\Phi_n^m 
= \frac{1}{\sqrt{-1}}\log\left(\frac{\tau_{n}^{m+1}\tau_n^m}{\taus_{n}^{m+1}\taus_n^m}\right)
=\frac{\Theta_{n}^{m+1}+\Theta_{n}^m}{2},\label{segment_m}
\end{equation}
which yields 
\begin{equation}
 \frac{\gamma_{n}^{m+1} -\gamma_n^m}{b_m}=R(\Phi_n^m-\Psi_n^m)~\frac{\gamma_{n+1}^m
 -\gamma_n^m}{a_n},\quad \Phi_n^m-\Psi_n^m=\frac{\Theta_n^{m+1}-\Theta_{n+1}^m}{2}=W_n^m.
\end{equation}
This is equivalent to equation (\ref{gamma_m}).

Finally let us derive the discrete potential mKdV equation (\ref{eqn:dpmKdV}). 
Dividing equation (\ref{bl6}) and its complex conjugate by
$\taus_{n+1}^m\taus_n^{m+1}$ we have
\begin{equation}
\begin{array}{l}\medskip
 {\displaystyle  b_m\exp\left(\frac{\sqrt{-1}~\Theta^{m}_{n+1}}{2}\right) 
- a_n\exp\left(\frac{\sqrt{-1}~\Theta^{m+1}_{n}}{2}\right) 
= - (a_n-b_m)\frac{\tau^*{}^{m+1}_{n+1}\tau^{m}_n}{\tau^*{}^{m+1}_n\tau^*{}^m_{n+1}},}\\
{\displaystyle b_m\exp\left(\frac{\sqrt{-1}~\Theta^{m+1}_{n}}{2}\right)  
- a_n\exp\left(\frac{\sqrt{-1}~\Theta^{m}_{n+1}}{2}\right) 
= -  (a_n-b_m)\frac{\tau^{m+1}_{n+1}\tau^*{}^{m}_n}{\tau^*{}^{m+1}_n\tau^*{}^m_{n+1}}},
\end{array}
\end{equation}
respectively. Dividing these two equations we obtain
\begin{equation}
\frac{b_m\exp\left(\frac{\sqrt{-1}~\Theta^{m}_{n+1}}{2}\right) 
- a_n\exp\left(\frac{\sqrt{-1}~\Theta^{m+1}_{n}}{2}\right) }
{b_m\exp\left(\frac{\sqrt{-1}~\Theta^{m+1}_{n}}{2}\right)  
- a_n\exp\left(\frac{\sqrt{-1}~\Theta^{m}_{n+1}}{2}\right)}
=
\exp\left[-\frac{\sqrt{-1}~\left(\Theta^{m+1}_{n+1} - \Theta^{m}_{n}\right)}{2}\right],
\label{th1:proof2}
\end{equation}
which is easily verified to be equivalent to equation (\ref{eqn:dpmKdV}).
Thus we have completed the proof of Theorem \ref{thm:tau_formula}.
\qquad$\square$

\begin{corollary}{\rm (\textbf{Representation formula})}\\
The function $\gamma_n^m$ can be expressed in terms of the potential function
 $\Theta_n^m$ as follows:
\begin{equation}
 \gamma_n^m(x,t;y)=\left[
\begin{array}{l}\medskip
 {\displaystyle \int^x \cos\Theta_n^m(x',t;y)~dx'}\\
 {\displaystyle \int^x \sin\Theta_n^m(x',t;y)~dx'}
\end{array}
\right]
=\left[
\begin{array}{l}\medskip
 {\displaystyle \sum^{n-1}_{n'} a_{n'}\cos\left(\frac{\Theta_{n'}^m(x,t;y)+\Theta_{n'+1}^m(x,t;y)}{2}\right)}\\
 {\displaystyle \sum^{n-1}_{n'} a_{n'}\sin\left(\frac{\Theta_{n'}^m(x,t;y)+\Theta_{n'+1}^m(x,t;y)}{2}\right)}
\end{array}
\right].
\end{equation}
\end{corollary}
\noindent\textbf{Proof.} The first equation is a consequence of
\begin{equation}
 \frac{\partial}{\partial x}\gamma_n^m(x,t;y) = \left[\begin{array}{c}\smallskip\cos\Theta_n^m(x,t;y)\\ \sin\Theta_n^m(x,t;y) \end{array}\right],
\end{equation}
and the second equation follows from equation (\ref{segment_n}).\qquad $\square$
\vskip5mm
It should be noted here that the bilinear equations (\ref{bl1})--(\ref{bl6}) are derived
from the reduction of the equations
\begin{align}
 & \frac{1}{2}D_xD_y~\tau_n^m(s)\cdot\tau_n^m(s) = -\tau_n^m(s+1)\tau_n^m(s-1),
\label{bl1_before_reduction}\\
 & \left(D_x^2-D_z\right)~\tau_n^m(s+1)\cdot\tau_n^m(s) =0,\label{bl2_before_reduction}\\
 &\left(D_x^3+D_t+3D_xD_z\right)~\tau_n^m(s+1)\cdot\tau_n^m(s) = 0,\label{bl3_before_reduction}\\
 & D_y~\tau_{n+1}^m(s)\cdot\tau_n^m(s) 
= -a_n\tau_{n+1}^m(s+1)\tau_n^m(s-1),\label{bl4_before_reduction}\\
 & D_y~\tau_{n}^{m+1}(s)\cdot\tau_n^m(s)
= -b_m\tau_{n+1}^m(s+1)\tau_n^m(s-1),\label{bl5_before_reduction}\\
 &b_m\tau_{n}^{m+1}(s+1)\tau_{n+1}^m(s) - a_n\tau_{n+1}^{m}(s+1)\tau_{n}^{m+1}(s)
+ (a_n-b_m)\tau_{n+1}^{m+1}(s+1)\tau_n^m(s)=0,\label{bl6_before_reduction}
\end{align}
for $\tau_n^m(s)=\tau_n^m(x,z,t;y;s)$, which are included in the discrete two-dimensional Toda
lattice hierarchy\cite{Jimbo-Miwa,Tsujimoto,Ueno-Takasaki}. In fact, imposing the conditions
\begin{equation}
\frac{\partial}{\partial z}\tau_n^m(s)=B~\tau_n^m(s),\quad  \tau_n^m(s+1) =C~\tau^*{}_n^m(s),\quad
B,C\in\mathbb{R},
\label{reduction_condition}
\end{equation}
and denoting $\tau_n^m=\tau_n^m(0)$, then
equations (\ref{bl1_before_reduction})--(\ref{bl6_before_reduction}) yield equations (\ref{bl1})--(\ref{bl6}),
respectively.
%
\section{B\"acklund transformations}
We start with the following fundamental fact on plane curves.
\begin{proposition}\label{prop:BTx}
Let $\gamma(x)$  be an arc-length parametrized curve with angle function $\theta(x)$.
Take a non-zero constant $\lambda$ and a solution $\widetilde\theta(x)$ to 
\begin{equation}
\left(\frac{\widetilde\theta +\theta}{2}\right)'
=2\lambda\sin\frac{\widetilde\theta-\theta}{2}.\label{BT3:pmKdV}
\end{equation}
Then 
\begin{equation}
 \widetilde{\gamma}(x)=\gamma(x)+\frac{1}{\lambda}R~\left(\frac{\widetilde\theta(x)-\theta(x)}{2}\right)~\gamma'(x)
\label{BTx:gamma}
\end{equation}
is an arc-length parametrized curve with angle function $\widetilde\theta(x)$. In other words, if
 $\gamma(x)$ is a solution to equation (\ref{gamma_x}), then $\widetilde\gamma(x)$ is another solution to
 equation (\ref{gamma_x}) with $\widetilde\kappa(x)=\widetilde\theta'(x)$.
The curve
 $\widetilde{\gamma}$ is called a B\"acklund transform of $\gamma$.
\end{proposition}
Proposition \ref{prop:BTx} can be verified easily by direct computation. 
We next extend the B\"acklund transformation to those of the motion of a curve. 
%
\begin{proposition}\label{prop:BT_continuous}
 Let $\gamma(x,t)$ be a motion of an arc-length parametrized curve determined by equations (\ref{gamma_t})
 and (\ref{eqn:pmKdV}). Take a B\"acklund transform $\widetilde\theta(x,t)$ defined by
 equations (\ref{BT1:pmKdV}) and (\ref{BT2:pmKdV}) of $\theta(x,t)$. Then 
\begin{equation}
  \widetilde{\gamma}(x,t)=\gamma(x,t)+\frac{1}{\lambda}R~\left(\frac{\widetilde\theta(x,t)-\theta(x,t)}{2}\right)~\gamma'(x,t)
\label{BT_continuous}
\end{equation}
is a motion of an arc-length parametrized curve with the angle function $\widetilde\theta(x,t)$.
\end{proposition}
\noindent\textbf{Proof.} By the preceding proposition, $\widetilde\gamma$ satisfies the
isoperimetric condition $|\widetilde\gamma'|=1$. Computing the $t$-derivative of
$\widetilde\gamma$ by using (\ref{BT2:pmKdV}), we can show that $\widetilde\gamma$ satisfies
equation (\ref{gamma_t}) with $\widetilde\kappa=\widetilde\theta'$\qquad $\square$\par\bigskip

%
Now we introduce a B\"acklund transformation of a discrete curve.
%
\begin{proposition}\label{prop:discrete_BT0}
 Let $\gamma_n$ be a discrete curve of segment length $a_n$. 
Let  $\Theta_n$ be the potential function  defined by 
\begin{equation}
 \frac{\gamma_{n+1}-\gamma_n}{a_n} 
= \left[\begin{array}{c}\cos\Psi_n \\\sin\Psi_n\end{array}\right],\quad
\Psi_n = \frac{\Theta_{n+1}+\Theta_{n}}{2}.\label{discrete_angle:2}
\end{equation}
For a non-zero constant $\lambda$, take a solution $\widetilde\Theta_n$ to the following equation:
\begin{equation}
  \tan\left(\frac{\widetilde\Theta_{n+1} - \Theta_n}{4}\right)=\frac{\frac{1}{\lambda}+a_n}{\frac{1}{\lambda}-a_n}\tan\left(\frac{\widetilde\Theta_n-\Theta_{n+1}}{4}\right).\label{BT0:dpmKdV}
\end{equation}
Then 
\begin{equation}
 \widetilde\gamma_n
=\gamma_n  +\frac{1}{\lambda}~R\left(\frac{\widetilde\Theta_n-\Theta_{n+1}}{2}\right)~
\frac{\gamma_{n+1}-\gamma_n}{a_n}\label{BTn:gamma}
\end{equation}
is a discrete curve with the potential function $\widetilde\Theta_n$.
\end{proposition}
%
\noindent\textbf{Proof.} It suffices to show that
\begin{equation}
  \frac{\widetilde\gamma_{n+1}-\widetilde\gamma_n}{a_n} 
= \left[\begin{array}{c}\cos\widetilde\Psi_n \\\sin\widetilde\Psi_n\end{array}\right],\quad
\widetilde\Psi_n = \frac{\widetilde\Theta_{n+1}+\widetilde\Theta_{n}}{2}\label{iso:gammatilde_n}
\end{equation}
for $\widetilde\gamma_n$ defined by equation (\ref{BTn:gamma}). This follows from
equations (\ref{discrete_angle:2}) and (\ref{BT0:dpmKdV}).\quad $\square$
\par\bigskip

We next extend the B\"acklund transformation to those of the motion of a discrete curve. In order to do
so, we first present the B\"acklund transformation to the discrete potential mKdV equation.
%
\begin{lemma}\label{lem:BT_dpmKdV}
Let $\Theta_n^m$ be a solution to the discrete potential mKdV equation (\ref{eqn:dpmKdV}). A
 function $\widetilde\Theta_n^m$ satisfying the following system of equations
\begin{align}
&  \tan\left(\frac{\widetilde\Theta_{n+1}^m - \Theta_n^m}{4}\right)=\frac{\frac{1}{\lambda}+a_n}{\frac{1}{\lambda}-a_n}\tan\left(\frac{\widetilde\Theta_n^m-\Theta_{n+1}^m}{4}\right),\label{BT1:dpmKdV}\\[2mm]
&  \tan\left(\frac{\widetilde\Theta_n^{m+1} - \Theta_n^m}{4}\right)=\frac{\frac{1}{\lambda}+b_m}{\frac{1}{\lambda}-b_m}\tan\left(\frac{\widetilde\Theta_n^m-\Theta_{n}^{m+1}}{4}\right),\label{BT2:dpmKdV}
\end{align}
gives another solution to equation (\ref{eqn:dpmKdV}). We call $\widetilde\Theta_n^m$ a B\"acklund
 transform of $\Theta_n^m$.
\end{lemma}
\noindent\textbf{Proof.} First note that equation (\ref{eqn:dpmKdV}) is equivalent to
\begin{equation}
 e^{U_n^{m+1}+U_n^m} - e^{U_{n+1}^{m+1}+U_{n+1}^m} = \frac{a_n}{b_m}\left( e^{U_{n+1}^{m}+U_n^m} - e^{U_{n+1}^{m+1}+U_{n}^{m+1}}\right),
\label{lem4:eq1}
\end{equation}
where we put $\frac{\sqrt{-1}\Theta_n^m}{2}=U_n^m$ for notational simplicity. Similarly, equations (\ref{BT1:dpmKdV}) and
(\ref{BT2:dpmKdV}) are rewritten as
\begin{align}
& e^{\widetilde{U}_n^{m}+U_n^m} - e^{\widetilde{U}_{n+1}^{m}+U_{n+1}^m} 
= \lambda a_n\left( e^{U_{n+1}^{m}+U_n^m} - e^{\widetilde{U}_{n+1}^{m}+\widetilde{U}_{n}^{m}}\right), \label{lem4:eq2}\\
& e^{\widetilde{U}_n^{m}+U_n^m} - e^{\widetilde{U}_{n}^{m+1}+U_{n}^{m+1}} 
= \lambda b_m\left( e^{U_{n}^{m+1}+U_n^m} - e^{\widetilde{U}_{n}^{m+1}+\widetilde{U}_{n}^{m}}\right) ,\label{lem4:eq3}
\end{align}
respectively, where $\frac{\sqrt{-1}\widetilde\Theta_n^m}{2}=\widetilde{U}_n^m$. Subtracting
equation (\ref{lem4:eq3}) from equation (\ref{lem4:eq2}), we have
\begin{equation}
e^{\widetilde{U}_{n}^{m+1}+U_{n}^{m+1}} -  e^{\widetilde{U}_{n+1}^{m}+U_{n+1}^m} 
= \lambda\left(  a_n e^{U_{n+1}^{m}+U_n^m} - b_m e^{U_{n}^{m+1}+U_n^m} \right)
- \lambda\left(  a_n e^{\widetilde{U}_{n+1}^{m}+\widetilde{U}_{n}^{m}} - b_me^{\widetilde{U}_{n}^{m+1}+\widetilde{U}_{n}^{m}}\right).\label{lem4:eq4}
\end{equation}
Similarly, subtracting equation (\ref{lem4:eq3})${}_{n\to n+1}$ from equation (\ref{lem4:eq2})${}_{m\to m+1}$, we get
\begin{equation}
e^{\widetilde{U}_n^{m+1}+U_n^{m+1}} -   e^{\widetilde{U}_{n+1}^{m}+U_{n+1}^m}
= \lambda\left( a_n e^{U_{n+1}^{m+1}+U_n^{m+1}} - b_m e^{U_{n+1}^{m+1}+U_{n+1}^m} \right)
- \lambda\left( a_n e^{\widetilde{U}_{n+1}^{m+1}+\widetilde{U}_{n}^{m+1}} - b_m e^{\widetilde{U}_{n+1}^{m+1}+\widetilde{U}_{n+1}^{m}}\right).\label{lem4:eq5}
\end{equation}
Subtracting equation (\ref{lem4:eq5}) from equation (\ref{lem4:eq4}) yields
\begin{equation}
\begin{split}
&  a_n \left(e^{\widetilde{U}_{n+1}^{m}+\widetilde{U}_{n}^{m}} - e^{\widetilde{U}_{n+1}^{m+1}+\widetilde{U}_{n}^{m+1}}\right)
- b_m\left( e^{\widetilde{U}_{n}^{m+1}+\widetilde{U}_{n}^{m}} - e^{\widetilde{U}_{n+1}^{m+1}+\widetilde{U}_{n+1}^{m}}\right) \\
&=  a_n \left(e^{U_{n+1}^{m}+U_{n}^{m}} - e^{U_{n+1}^{m+1}+U_{n}^{m+1}}\right)- b_m\left( e^{U_{n}^{m+1}+U_{n}^{m}} - e^{U_{n+1}^{m+1}+U_{n+1}^{m}}\right).
\end{split}\label{lem4:eq6}
\end{equation}
Now we see that the right-hand side of equation (\ref{lem4:eq6}) vanishes since it is equivalent to
equation (\ref{lem4:eq1}). Then the left-hand side gives equation (\ref{eqn:dpmKdV}) for
$\widetilde\Theta_n^m$.\qquad $\square$.\par\bigskip
%
\begin{proposition}\label{prop:BT_discrete}
 Let $\gamma_n^m$ be a discrete motion of a discrete curve. Take a B\"acklund transform $\widetilde\Theta_n^m$ of
$\Theta_n^m$ defined in Lemma \ref{lem:BT_dpmKdV}. Then 
\begin{equation}
  \widetilde\gamma_n^m = \gamma_n^m + \frac{1}{\lambda}~
R\left(\frac{\widetilde\Theta_n^m -\Theta_{n+1}^m}{2}\right)
~\frac{\gamma_{n+1}^m-\gamma_n^m}{a_n}
\label{BT_discrete}
\end{equation}
is a discrete motion of a discrete curve with potential function $\widetilde\Theta_n^m.$ We call
 $\widetilde{\gamma}_n^m$ a B\"acklund transform of $\gamma_n^m$.
\end{proposition}
%
\noindent\textbf{Proof.} It suffices to show that $\widetilde\gamma_n^m$ satisfies
equations (\ref{iso})--(\ref{gamma_m}) with potential function $\widetilde\Theta_n^m$. But equations
(\ref{iso}) and (\ref{gamma_n}) follow from Proposition \ref{prop:discrete_BT0} immediately.  Noticing the
symmetry in $n$ and $m$, similar calculations to those in Proposition \ref{prop:discrete_BT0} yield
\begin{equation}
  \frac{\widetilde\gamma_{n}^{m+1}-\widetilde\gamma_n^m}{b_m} 
=\left[\begin{array}{c}\smallskip\cos\left(\frac{\widetilde\Theta_{n}^{m+1}+\widetilde\Theta_{n}^m}{2}\right) \\
      \sin\left(\frac{\widetilde\Theta_{n}^{m+1}+\widetilde\Theta_{n}^m}{2}\right)\end{array}\right]
\label{iso:gammatilde_m}
\end{equation}
by using equation (\ref{BT2:dpmKdV}). Comparing equations (\ref{iso:gammatilde_n}) and (\ref{iso:gammatilde_m})
we obtain 
\begin{equation}
  \frac{\widetilde\gamma_{n}^{m+1}-\widetilde\gamma_n^m}{b_m} =R\left(\frac{\widetilde\Theta_n^{m+1}-\widetilde\Theta_{n+1}^m}{2}\right)~
 \frac{\widetilde\gamma_{n+1}^{m}-\widetilde\gamma_n^m}{a_n},
\end{equation}
which implies equation (\ref{gamma_m}).\qquad $\square$
\par\bigskip
It is possible to construct another type of B\"acklund transformation for motions of both
smooth and discrete curves by using the symmetry of the potential mKdV equation
(\ref{eqn:pmKdV}) and discrete potential mKdV equation (\ref{eqn:dpmKdV}). In fact, if
$\theta(x,t)$ is a solution to equation (\ref{eqn:pmKdV}), then $-\theta(x,t)$ satisfies the same
equation. Combining this symmetry and the B\"acklund transformation defined by
equations (\ref{BT1:pmKdV}) and (\ref{BT2:pmKdV}), we have the following B\"acklund transformation.
%
\begin{lemma}\label{lem:BT2_pmKdV} 
Let $\theta(x,t)$ be a solution to the potential mKdV equation (\ref{eqn:pmKdV}). For any non-zero constant
$\lambda$, a function $\overline\theta(x,t)$ satisfying the following set of equations
\begin{align}
&\frac{\partial}{\partial x}\left(\frac{\overline\theta -\theta}{2}\right)
=2\lambda\sin\frac{\overline\theta+\theta}{2},\label{BT4:pmKdV}\\[1mm]
 &\frac{\partial}{\partial t}\left(\frac{\overline\theta - \theta}{2}\right)
=-\lambda\left\{\left(\theta_x\right)^2+8\lambda^2\right\}
\sin\frac{\overline\theta + \theta}{2}
-2\lambda\theta_{xx}\cos\frac{\overline\theta+\theta}{2}
-4\lambda^2\theta_x,\label{BT5:pmKdV}
\end{align}
gives another solution to equation (\ref{eqn:pmKdV}).
\end{lemma}
Lemma \ref{lem:BT2_pmKdV} immediately yields the following B\"acklund transformation for $\gamma(x,t)$.
%
\begin{proposition}\label{prop:BT2_continuous}
Let $\gamma(x,t)$ be a motion of an arc-length parametrized curve determined by equations (\ref{gamma_t})
 and (\ref{eqn:pmKdV}). Take a B\"acklund transform $\overline\theta(x,t)$ of
$\theta(x,t)$ defined in Lemma \ref{lem:BT2_pmKdV}. Then 
\begin{equation}
 \overline\gamma(x,t) = S\left[\gamma(x,t) + \frac{1}{\lambda}~R\left(-\frac{\overline\theta(x,t) +\theta(x,t)}{2}\right)
~\gamma'(x,t)\right],\quad S=\left[\begin{array}{cc}1 &0 \\ 0 &-1 \end{array}\right],\label{BT2:gamma}
\end{equation}
is a motion of an arc-length parametrized curve with angle function $\overline\theta(x,t)$.
 \end{proposition}
%
Note that equations (\ref{BT4:pmKdV}) and (\ref{BT5:pmKdV}) can be derived from equations (\ref{BT1:pmKdV}) and
(\ref{BT2:pmKdV}) simply by putting
$\widetilde\theta(x,t)=-\overline{\theta}(x,t)$. Moreover, putting
\begin{displaymath}
\hat\gamma(x,t):= \gamma(x,t) + \frac{1}{\lambda}~R\left(-\frac{\overline\theta(x,t) +\theta(x,t)}{2}\right)
~\gamma'(x,t),
\end{displaymath}
and noticing equation (\ref{angle_function}) and Proposition
\ref{prop:BT_continuous}, we have
\begin{equation}
\hat\gamma'(x,t) = 
\left[\begin{array}{c}\cos(-\overline\theta(x,t))\\\sin(-\overline\theta(x,t))\end{array}\right]
=\left[\begin{array}{c}\cos\overline\theta(x,t)\\-\sin\overline\theta(x,t)\end{array}\right],
\end{equation}
which implies Proposition \ref{prop:BT2_continuous}.

Similarly, if $\Theta_n^m$ is a solution to equation (\ref{eqn:dpmKdV}), then $-\Theta_n^m$ satisfies the
same equation. Therefore Lemma \ref{lem:BT_dpmKdV} and Proposition \ref{prop:BT_discrete} lead to
the following B\"acklund transformations.
%
\begin{lemma} \label{lem:BT2_dpmKdV}
Let $\Theta_n^m$ be a solution to the discrete potential mKdV equation (\ref{eqn:dpmKdV}). A function
 $\overline\Theta{}_n^m$ satisfying the following system of equations
\begin{align}
&  \tan\left(\frac{\overline\Theta{}_{n+1}^m + \Theta_n^m}{4}\right)=\frac{\frac{1}{\lambda}+a_n}{\frac{1}{\lambda}-a_n}\tan\left(\frac{\overline\Theta{}_n^m+\Theta_{n+1}^m}{4}\right),\label{BT3:dpmKdV}\\[2mm]
&  \tan\left(\frac{\overline\Theta{}_n^{m+1} + \Theta_n^m}{4}\right)=\frac{\frac{1}{\lambda}+b_m}{\frac{1}{\lambda}-b_m}\tan\left(\frac{\overline\Theta{}_n^m+\Theta_{n}^{m+1}}{4}\right),\label{BT4:dpmKdV}
\end{align}
gives another solution to equation (\ref{eqn:dpmKdV}). 
\end{lemma}
%
\begin{proposition} \label{prop:BT2_discrete}
 Let $\gamma_n^m$ be a discrete motion of a discrete curve. Take a B\"acklund transform $\overline\Theta{}_n^m$ of
$\Theta_n^m$ defined in Lemma \ref{lem:BT2_dpmKdV}. Then 
\begin{equation}
 \overline\gamma{}_n^m = S\left[\gamma_n^m + \frac{1}{\lambda}~R\left(-\frac{\overline\Theta{}_n^m +\Theta_{n+1}^m}{2}\right)
~\frac{\gamma_{n+1}^m-\gamma_n^m}{a_n}\right],\quad S=\left[\begin{array}{cc}1 &0 \\ 0 &-1 \end{array}\right],\label{BT2:gamma_discrete}
\end{equation}
is a discrete motion of a discrete curve with potential function $\overline\Theta{}_n^m.$ 
 \end{proposition}
\begin{remark}\rm\hfill
\begin{enumerate}
 \item It may be interesting to point out that equation (\ref{BT3:dpmKdV}) and equation (\ref{BT4:dpmKdV})
       can be rewritten as
 \begin{align}
&  \sin\left(\frac{\overline\Theta{}_{n+1}^m - \Theta_{n+1}^m - \overline\Theta{}_{n}^m + \Theta_{n}^m}{4}\right)
=\lambda a_n
\sin\left(\frac{\overline\Theta{}_{n+1}^m + \Theta_{n+1}^m + \overline\Theta{}_{n}^m + \Theta_{n}^m}{4}\right),
\label{BT5:dpmKdV}\\[2mm]
&  \sin\left(\frac{\overline\Theta{}_{n}^{m+1} - \Theta_{n}^{m+1} - \overline\Theta{}_{n}^m + \Theta_{n}^m}{4}\right)
=\lambda b_m
\sin\left(\frac{\overline\Theta{}_{n}^{m+1} + \Theta_{n}^{m+1} + \overline\Theta{}_{n}^m + \Theta_{n}^m}{4}\right),
\label{BT6:dpmKdV}
\end{align}
respectively, which are essentially equivalent to the discrete sine-Gordon equation\cite{Hirota:dsG}.
 \item The B\"acklund transformations described in Propositions \ref{prop:BT_continuous} and
\ref{prop:BT_discrete} satisfy the `constant distance property', i.e.,
$|\widetilde\gamma-\gamma|\equiv 1/\lambda$ or 
$|\widetilde\gamma_n^m-\gamma_n^m|\equiv 1/\lambda$. These transformations may be regarded as
the one-dimensional analogue of the original B\"acklund transformations of the pseudospherical
surface\cite{Rogers-Schief}. On the other hand, the B\"acklund transformations proposed in
Propositions \ref{prop:BT2_continuous} and \ref{prop:BT2_discrete} are characterized by the property $|\overline{\gamma} -S\gamma|=1/\lambda$.
\end{enumerate}
\end{remark}
\section{Explicit solutions}
\subsection{Solitons and breathers}
For $N\in\mathbb{Z}_{\geq 0}$ we define a function
$\tau_n^m(s)=\tau_n^m(x,t;y,z;s)$ by
\begin{equation}
 \tau_n^m(s) = \exp\left[-\left(x + \sum_{n'}^{n-1}a_{n'} + \sum_{m'}^{m-1}
b_{m'}\right)y\right]
~\det\left(f_{s+j-1}^{(i)}\right)_{i,j=1,\ldots,N},\label{Casorati:without_reduction}
\end{equation}
for $(x,t;y,z)\in\mathbb{R}^4$ and $(m,n,s)\in\mathbb{Z}^3$.
Here $f_{s}^{(i)}=f_{s}^{(i)}(x,t;y,z;m,n)$ ($i=1,\ldots,N$) satisfies the following linear
equations:
\begin{equation}
\frac{\partial f_s^{(i)}}{\partial x}=f_{s+1}^{(i)},\quad
\frac{\partial f_s^{(i)}}{\partial z}=f_{s+2}^{(i)},\quad
\frac{\partial f_s^{(i)}}{\partial t}=-4f_{s+3}^{(i)},\quad
\frac{\partial f_s^{(i)}}{\partial y}=f_{s-1}^{(i)},\label{Casorati:linear_continuous}
\end{equation}
\begin{equation}
 \frac{f_s^{(i)}(m,n)-f_s^{(i)}(m,n-1)}{a_{n-1}} = f_{s+1}^{(i)}(m,n),\quad
 \frac{f_s^{(i)}(m,n)-f_s^{(i)}(m-1,n)}{b_{m-1}} = f_{s+1}^{(i)}(m,n).\label{Casorati:linear_discrete}
\end{equation}
For $N=0$, we set $\det(f^{(i)}_{s+j-1})_{i,j=1,\ldots,N}=1$. 
A typical example for $f_s^{(i)}$ is given by
\begin{align}
&f_s^{(i)} = e^{\eta_i} + e^{\mu_i},\label{f_without_reduction}\\
&\left\{
\begin{array}{l}
 {\displaystyle  e^{\eta_i}= \alpha_ip_i^s \prod_{n'}^{n-1}(1-a_{n'}p_i)^{-1}\prod_{m'}^{m-1}(1-b_{m'}p_i)^{-1}
e^{p_i x + p_i^2 z - 4p_i^3 t +  \frac{1}{p_i}y }},\\[4mm]
{\displaystyle  e^{\mu_j}= \beta_iq_i^s \prod_{n'}^{n-1}(1-a_{n'}q_i)^{-1}\prod_{m'}^{m-1}(1-b_{m'}q_i)^{-1}
e^{q_i x  +  q_i^2 z - 4q_i^3 t +  \frac{1}{q_i}y }},
\end{array}\right.
\label{Casorati_entries:without_reduction}
\end{align}
where $p_i$,  $q_i$, $\alpha_i$ and $\beta_i$ are arbitrary complex constants.

We note that $\tau_n^m$ and $f_s^{(i)}$ are functions of continuous variables $x$, $y$, $z$, $t$ and
discrete variables $m$, $n$, $s$, but we will indicate only the relevant variables according to the
context, for notational simplicity. Then it is well known that $\tau_n^m(s)$ satisfies the bilinear
equations (\ref{bl1_before_reduction})--(\ref{bl6_before_reduction})
\cite{Hirota:book,Jimbo-Miwa,OHTI:dKP,OKMS:RT,MKO:dRT,MO:dNLS_dark,Ueno-Takasaki}. Actually by using
the linear relations (\ref{Casorati:linear_continuous}) and (\ref{Casorati:linear_discrete}),
equations (\ref{bl1_before_reduction})--(\ref{bl6_before_reduction}) are reduced to the Pl\"ucker
relations, which are quadratic identities of determinants.

It is possible to construct the solutions to the bilinear equations (\ref{bl1})--(\ref{bl6}) by
imposing the reduction condition (\ref{reduction_condition}) on $\tau_n^m(s)$ in
equation (\ref{Casorati:without_reduction}). Those conditions are realized by putting restrictions on
the parameters of the solutions. As an example, we present the multi-soliton and multi-breather
solutions.
\begin{proposition}\label{prop:soliton}
Consider the $\tau$ function
\begin{equation}
 \tau_n^m = \exp\left[-\left(x + \sum_{n'}^{n-1}a_{n'} + \sum_{m'}^{m-1}b_{m'}\right)y\right]
~\det\left(f_{j-1}^{(i)}\right)_{i,j=1,\ldots,N},\label{Casorati} 
\end{equation}
\begin{equation}
f_s^{(i)} = e^{\eta_i} + e^{\mu_i},\label{tau:Casorati_entries0}
\end{equation}
\begin{equation}
\left\{
\begin{array}{l}
 {\displaystyle  e^{\eta_i}= \alpha_ip_i^s \prod_{n'}^{n-1}(1-a_{n'}p_i)^{-1}\prod_{m'}^{m-1}(1-b_{m'}p_i)^{-1}
e^{p_i x - 4p_i^3 t +  \frac{1}{p_i}y }},\\[4mm]
{\displaystyle e^{\mu_j}=  \beta_i(-p_i)^s \prod_{n'}^{n-1}(1+a_{n'}p_i)^{-1}\prod_{m'}^{m-1}(1+b_{m'}p_i)^{-1}
e^{-p_ix  + 4p_i^3 t -  \frac{1}{p_i}y }}.
\end{array}\right.
\label{tau:Casorati_entries}
\end{equation}
\begin{enumerate}
 \item Choosing the parameters as
\begin{equation}\label{param:soliton}
p_i,\ \alpha_i\in\mathbb{R},\quad \beta_i\in\sqrt{-1}\mathbb{R}\quad (i=1,\ldots,N),
\end{equation}
then $\tau_n^m$ satisfies the bilinear equations (\ref{bl1})--(\ref{bl6}). This gives the
       $N$-soliton solution to equations  (\ref{eqn:pmKdV}) and (\ref{eqn:dpmKdV}).
 \item Taking $N=2M$, and choosing the parameters as
\begin{equation}\label{param:breather}
\begin{array}{l}\medskip
 {\displaystyle  p_i,\ \alpha_i,\ \beta_i \in\mathbb{C}\quad (i=1,\ldots,2M),
\quad p_{2k}=p_{2k-1}^*\quad (k=1,\ldots,M),}\\
{\displaystyle \alpha_{2k}=\alpha_{2k-1}^*,\quad  \beta_{2k}=-\beta_{2k-1}^*\quad (k=1,\ldots,M),}
\end{array}
\end{equation}
then $\tau_n^m$ satisfies the bilinear equations (\ref{bl1})--(\ref{bl6}). This gives the
       $M$-breather solution to equations  (\ref{eqn:pmKdV}) and (\ref{eqn:dpmKdV}).
\end{enumerate}
\end{proposition}
\noindent\textbf{Proof.} It is sufficient to show that the conditions in equation (\ref{reduction_condition}) are
satisfied. We first impose the two-periodicity in $s$, i.e., $\tau_n^m(s+2)={\rm const.}\times\tau_n^m(s)$. 
For $\tau_n^m(s)$ in equation (\ref{Casorati:without_reduction}) with entries given by
equations (\ref{f_without_reduction}) and (\ref{Casorati_entries:without_reduction}), putting
\begin{equation}
 q_i=-p_i,
\end{equation}
we have
\begin{equation}
 f_{s+2}^{(i)}=p_i^{2}f_s^{(i)},
\end{equation}
which implies that 
\begin{equation}
 \tau_n^m(s+2)=A_N~\tau_n^m(s),\quad A_N=\prod_{i=1}^N p_i^2.
\end{equation}
Note that the condition 
\begin{equation}
 \frac{\partial \tau_n^m(s)}{\partial z} = B_N~\tau_n^m(s),\quad B_N = \sum_{i=1}^N p_i^2,
\end{equation}
is also satisfied simultaneously.  Now we consider cases (1) and (2) separately.\par\bigskip

Case (1). We see from equations (\ref{tau:Casorati_entries0}) and (\ref{tau:Casorati_entries})
together with equation (\ref{param:soliton}) that
\begin{equation}
 f_{1}^{(i)} = p_i~f_0^{(i)}{}^*
\end{equation}
and so 
\begin{equation}
 \tau_n^m(1)=C_N~\taus_n^m(0),\quad C_N=\prod_{i=1}^N p_i\in\mathbb{R}.
\end{equation}

Case (2). We see from equations (\ref{tau:Casorati_entries0}) and (\ref{tau:Casorati_entries})
together with equation (\ref{param:breather}) that
\begin{equation}
 f_{1}^{(2k)} = p_{2k-1}^*~f_0^{(2k-1)}{}^*,\quad  f_{1}^{(2k-1)} = p_{2k}^*~f_0^{(2k)}{}^*,
\end{equation}
and so 
\begin{equation}
 \tau_n^m(1)=C_N~\taus_n^m(0),\quad C_N=(-1)^M\prod_{i=1}^M \left|p_{2i}\right|^2\in\mathbb{R}.
\end{equation}
Therefore we have verified that the conditions in equation (\ref{reduction_condition}) are satisfied for
both cases. Then putting $\tau_n^m=\tau_n^m(0)$, we obtain the desired result.\qquad
$\square$\par\bigskip

\begin{figure}[ht]
\begin{center}
\includegraphics[scale=0.5]{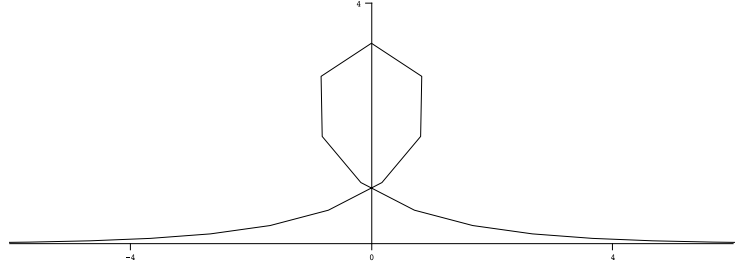} 
\caption{Parameters in equations (\ref{Casorati}),
(\ref{tau:Casorati_entries0}) and (\ref{tau:Casorati_entries}): $N=1$, 
$x=0$, $y=0$, $\alpha_1=-1$, 
$\beta_{1}=\sqrt{-1}$, $p_1=0.3$, $a_n=1$, $b_m=0.5$.}\label{fig:1-soliton}
\end{center}
\end{figure}

\begin{figure}[ht]
\begin{minipage}{.45\textwidth}
\begin{center}
\includegraphics[width=7truecm]{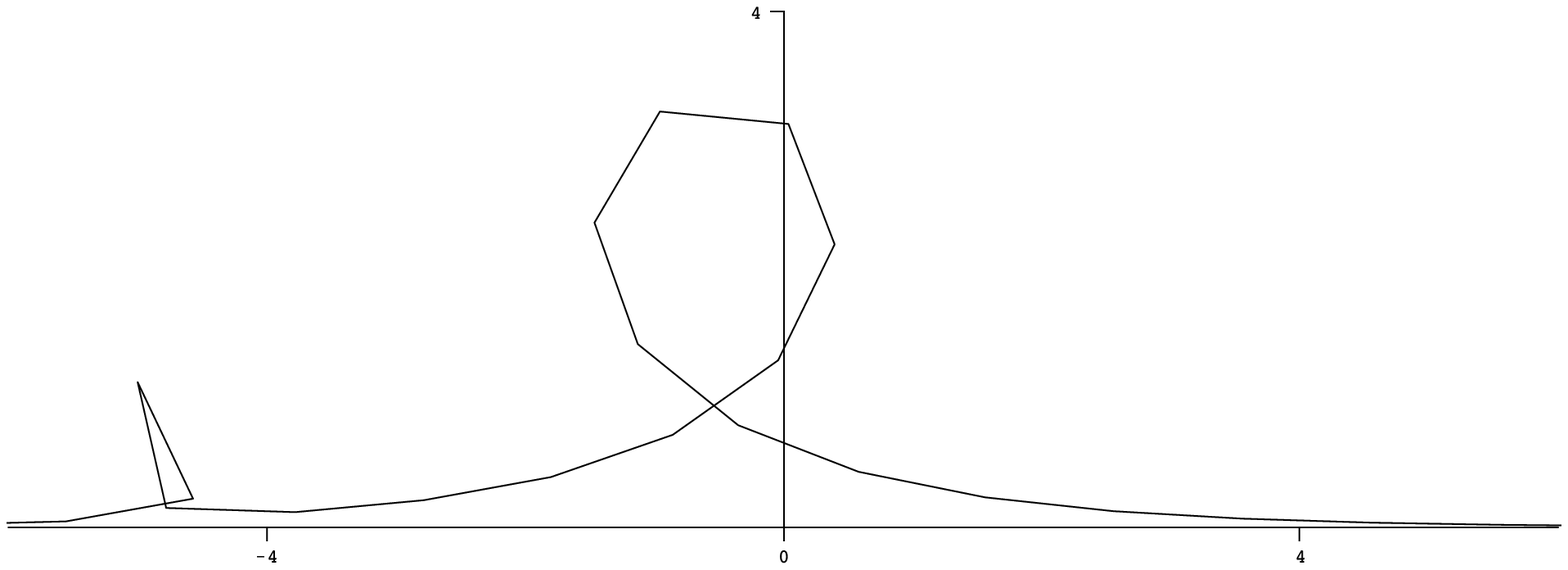}\\
$m = - 46$ 
\end{center}
\end{minipage}
\begin{minipage}{.45\textwidth}
\begin{center}
\includegraphics[width=7truecm]{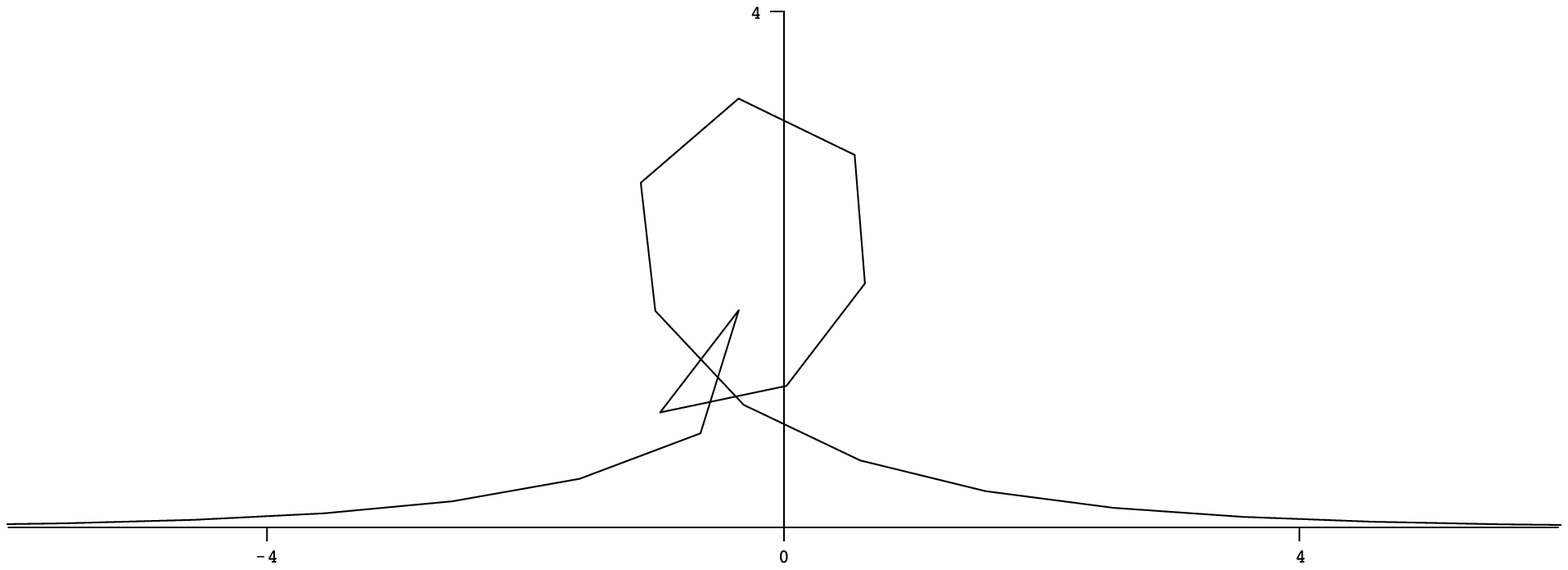}\\
$m = - 20$ 
\end{center}
\end{minipage}
\begin{minipage}{.45\textwidth}
\begin{center}
\includegraphics[width=7truecm]{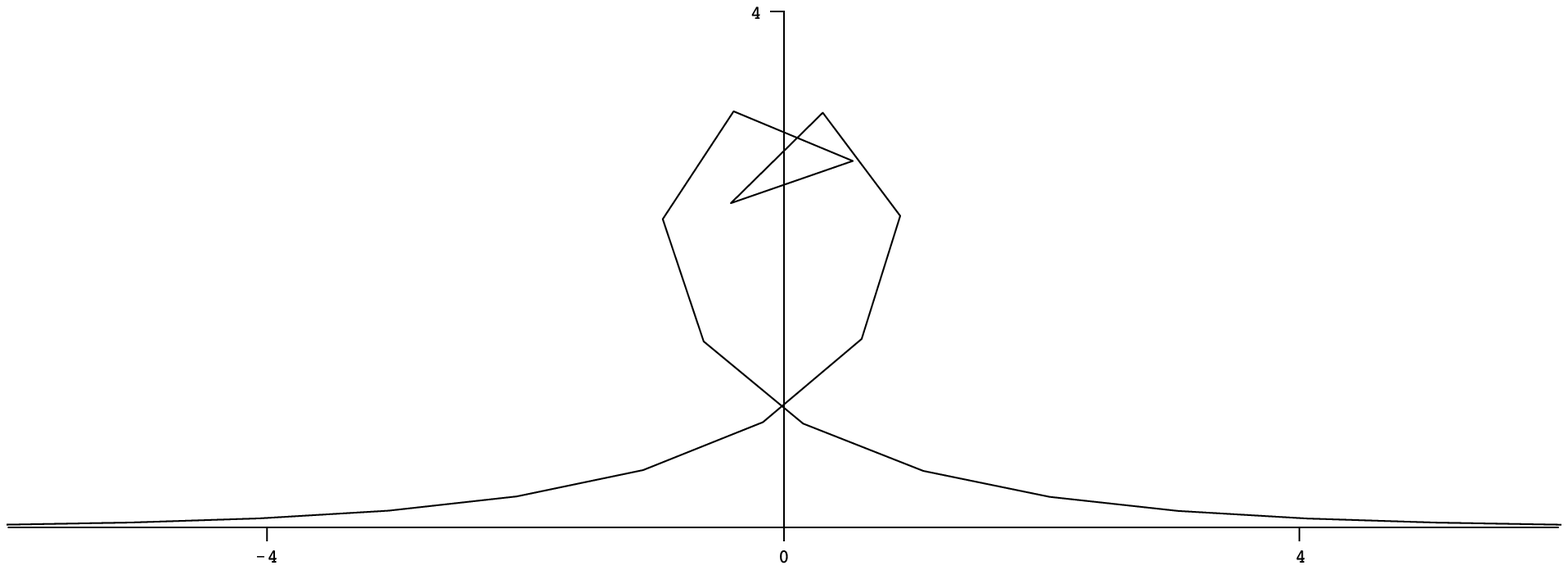}\\
$m = - 1$ 
\end{center}
\end{minipage}\hfill
\begin{minipage}{.45\textwidth}
\begin{center}
\includegraphics[width=7truecm]{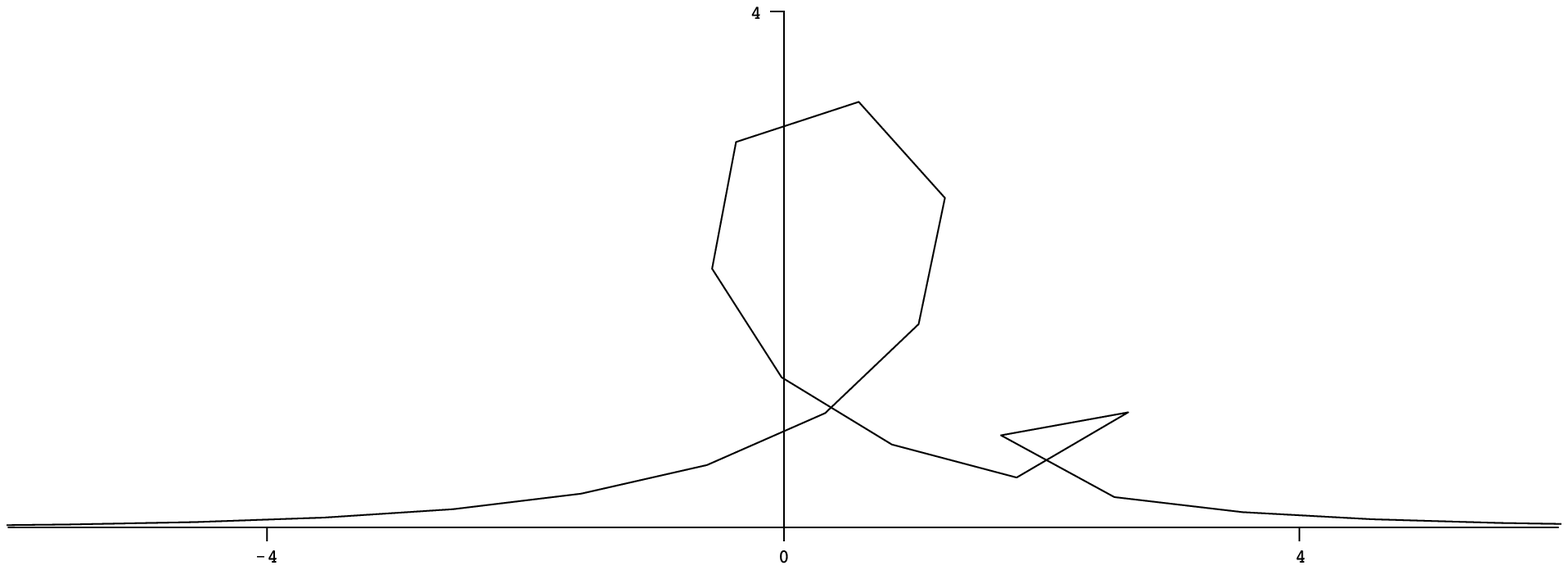}\\
$m = 30$ 
\end{center}
\end{minipage}
\caption{Parameters in equations (\ref{Casorati}),
(\ref{tau:Casorati_entries0}) and (\ref{tau:Casorati_entries}): 
$N=2$, $x=0$, $y=0$, $\alpha_{1}=-1$, $\alpha_{2}=1$, 
$\beta_{1}=-\beta_{2}=\sqrt{-1}$, $p_1=0.3$, $p_2=0.9$, $a_n=1$, $b_m=0.5$.}
\label{fig:2-soliton}
\end{figure}

\begin{figure}[ht]
\begin{minipage}{.45\textwidth}
\begin{center}
\includegraphics[width=7truecm]{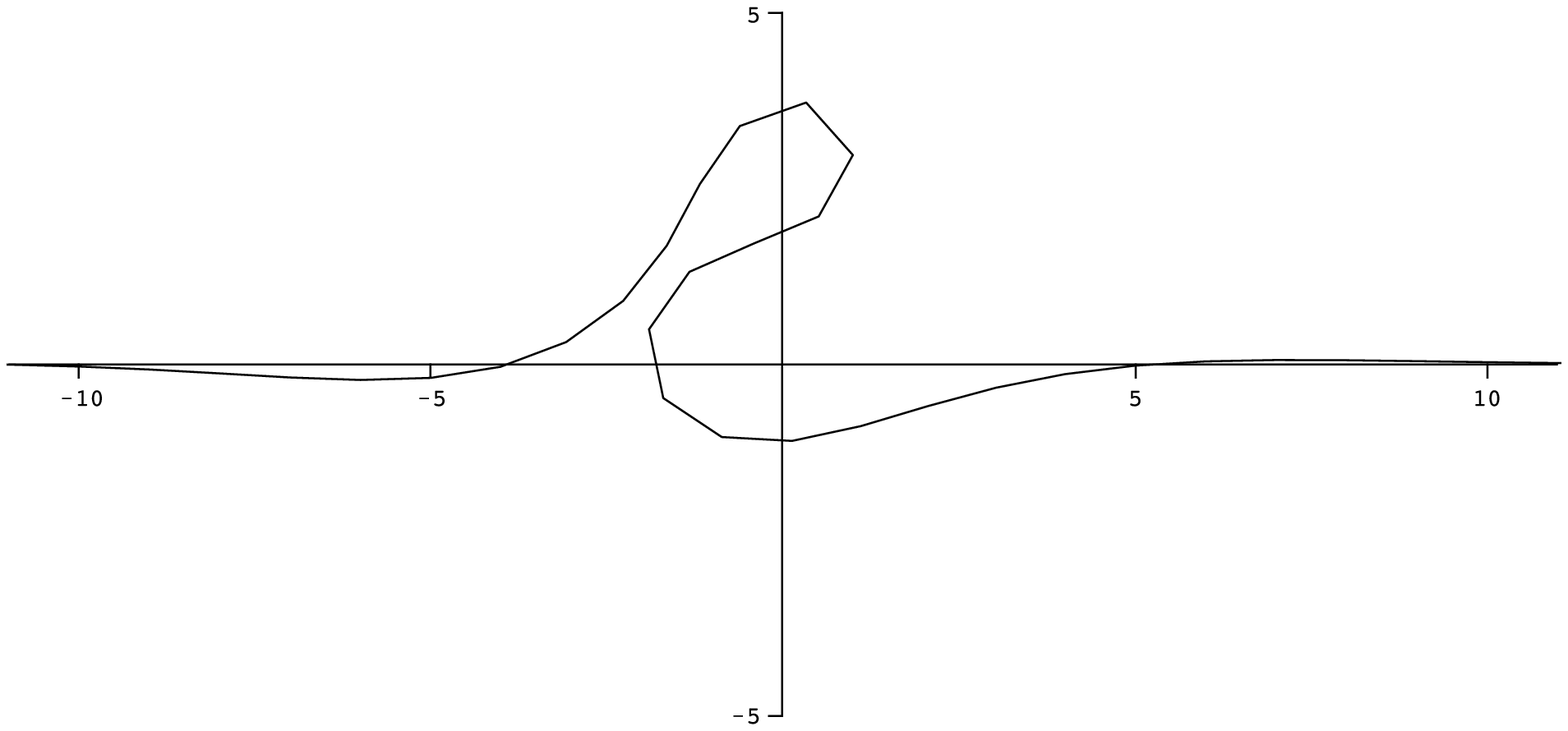}\\
$m = - 18$ 
\end{center}
\end{minipage}
\begin{minipage}{.45\textwidth}
\begin{center}
\includegraphics[width=7truecm]{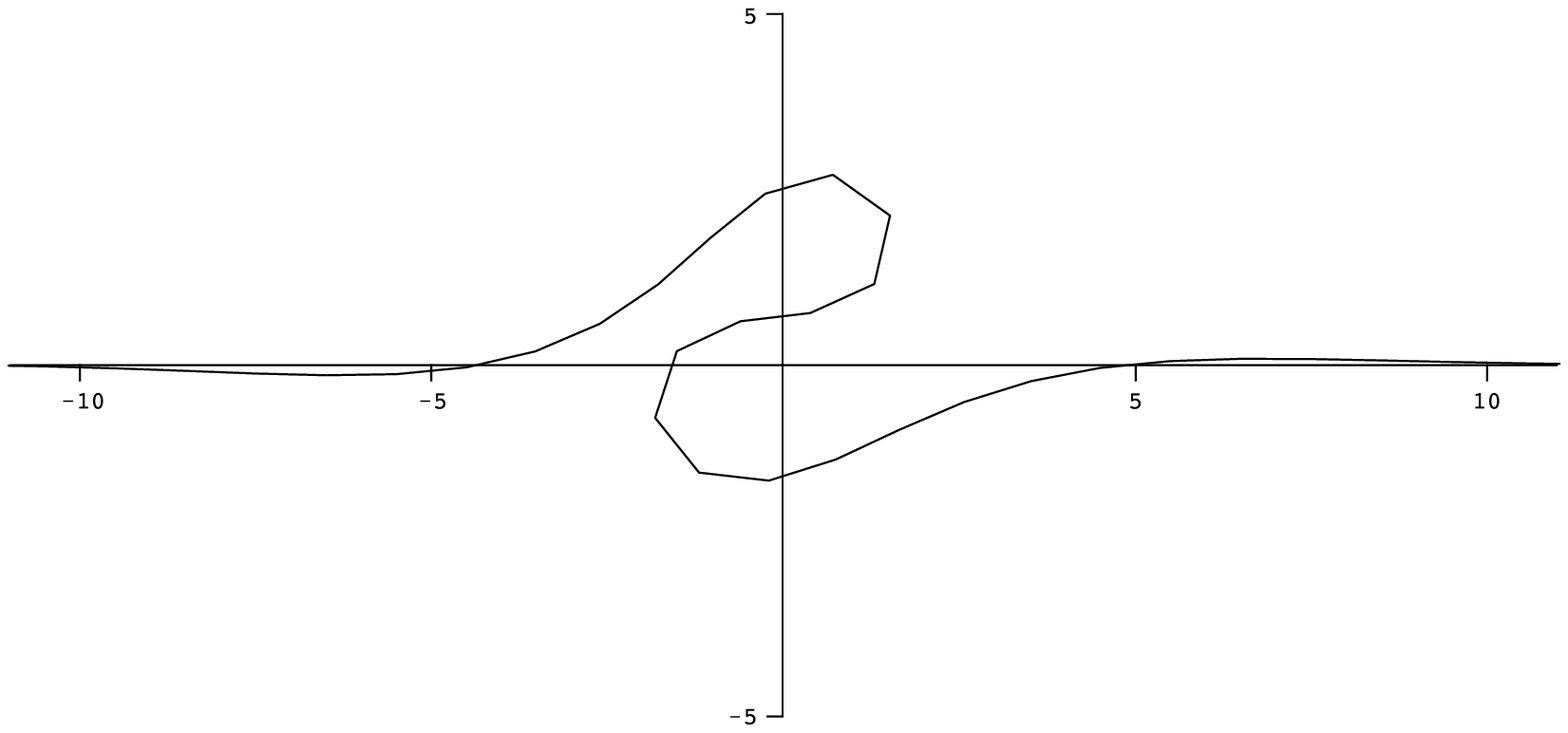}\\
$m = - 7$ 
\end{center}
\end{minipage}
\begin{minipage}{.45\textwidth}
\begin{center}
\includegraphics[width=7truecm]{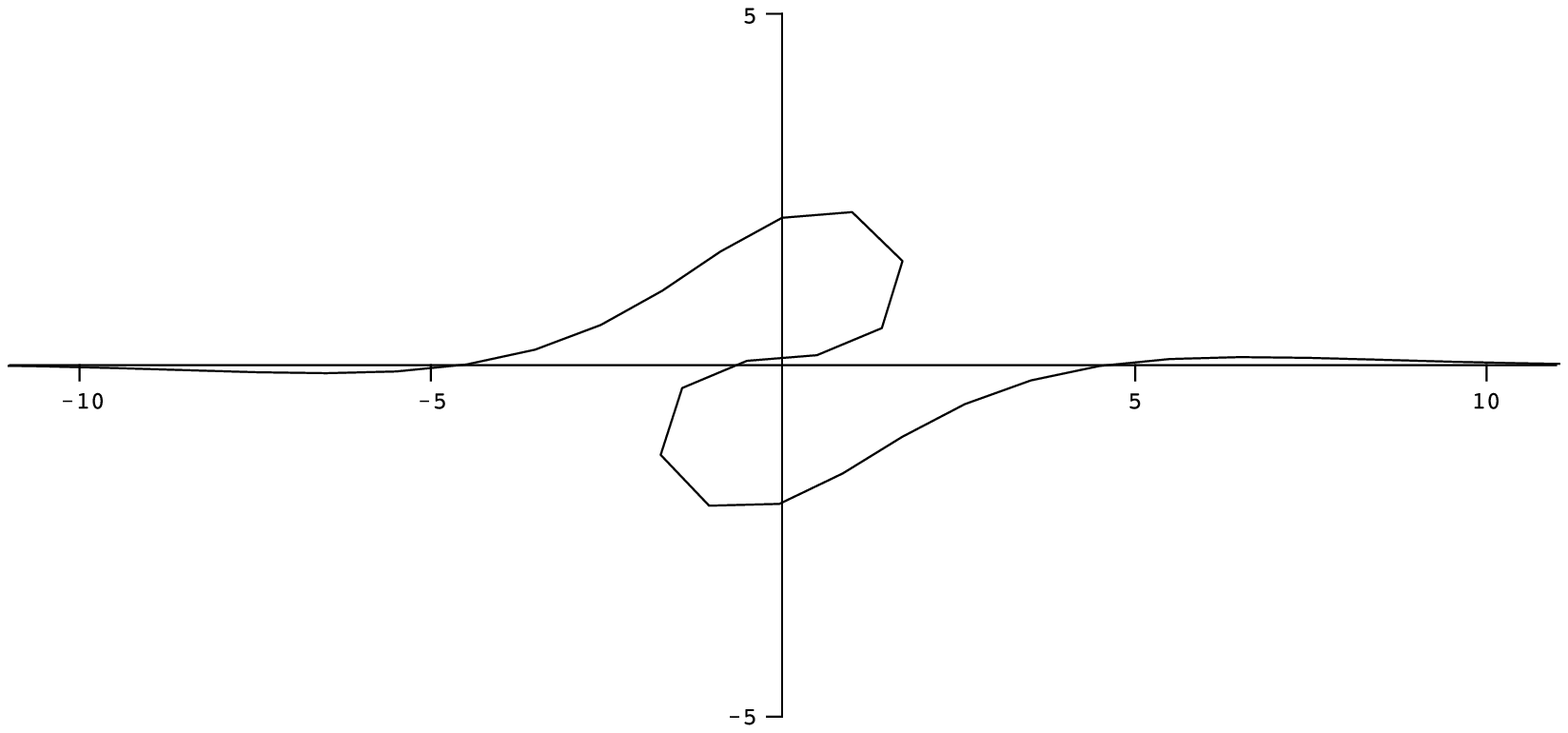}\\
$m = - 1$ 
\end{center}
\end{minipage}\hfill
\begin{minipage}{.45\textwidth}
\begin{center}
\includegraphics[width=7truecm]{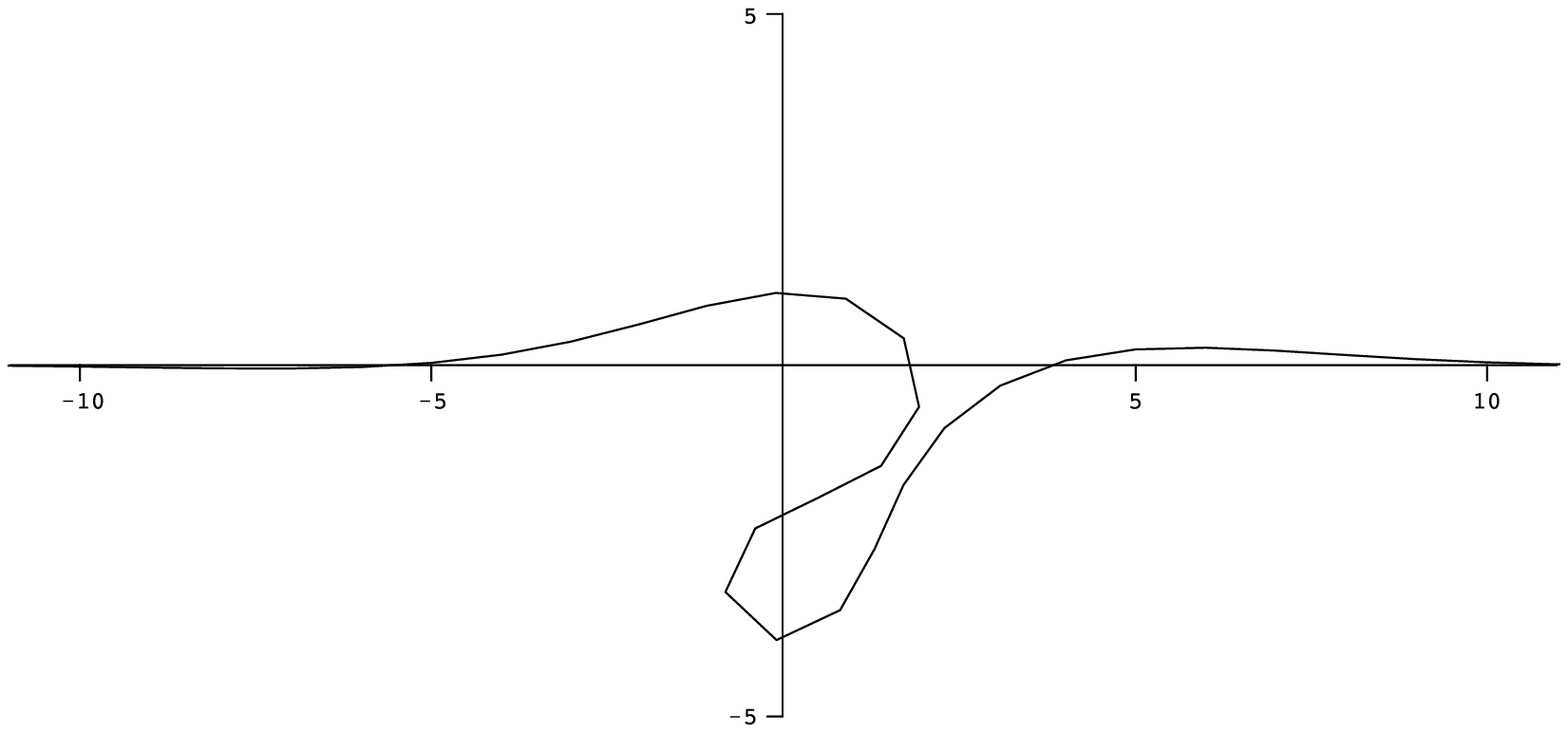}\\
$m = 20$ 
\end{center}
\end{minipage}
\caption{Parameters in equations (\ref{Casorati}), (\ref{tau:Casorati_entries0}) and
(\ref{tau:Casorati_entries}): $N=2$, $x=0$, $y=0$, 
$\alpha_{1}=\alpha_{2}^*=1$, 
$\beta_{1}=-\beta_{2}^*=1$, $p_1=p_2^*=0.2-0.2\sqrt{-1}$, $a_n=1$,
$b_m=1.5$.}\label{fig:1-breather}
\end{figure}

\begin{figure}[ht]
\begin{minipage}{.45\textwidth}
\begin{center}
\includegraphics[width=7truecm]{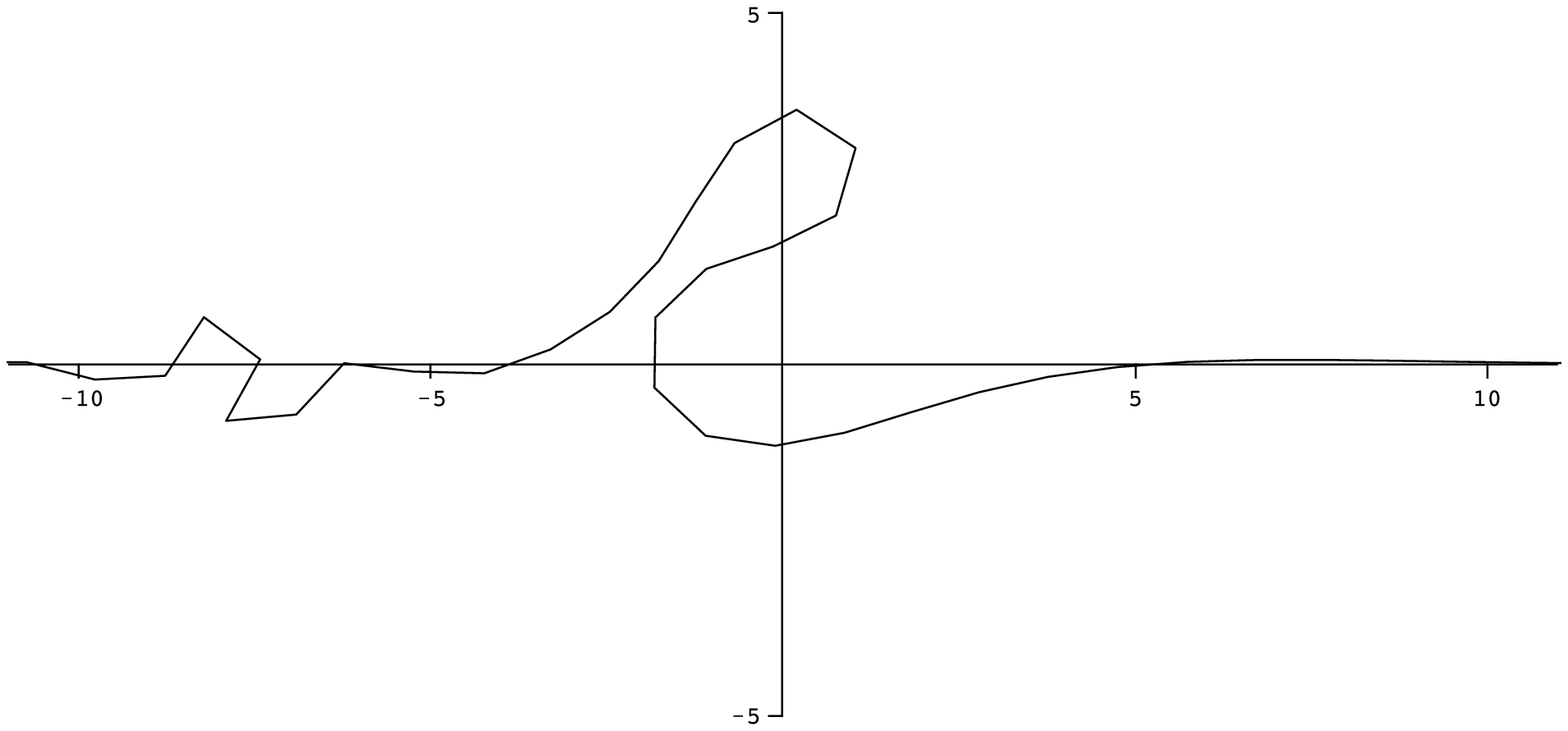}\\
$m = - 18$ 
\end{center}
\end{minipage}
\begin{minipage}{.45\textwidth}
\begin{center}
\includegraphics[width=7truecm]{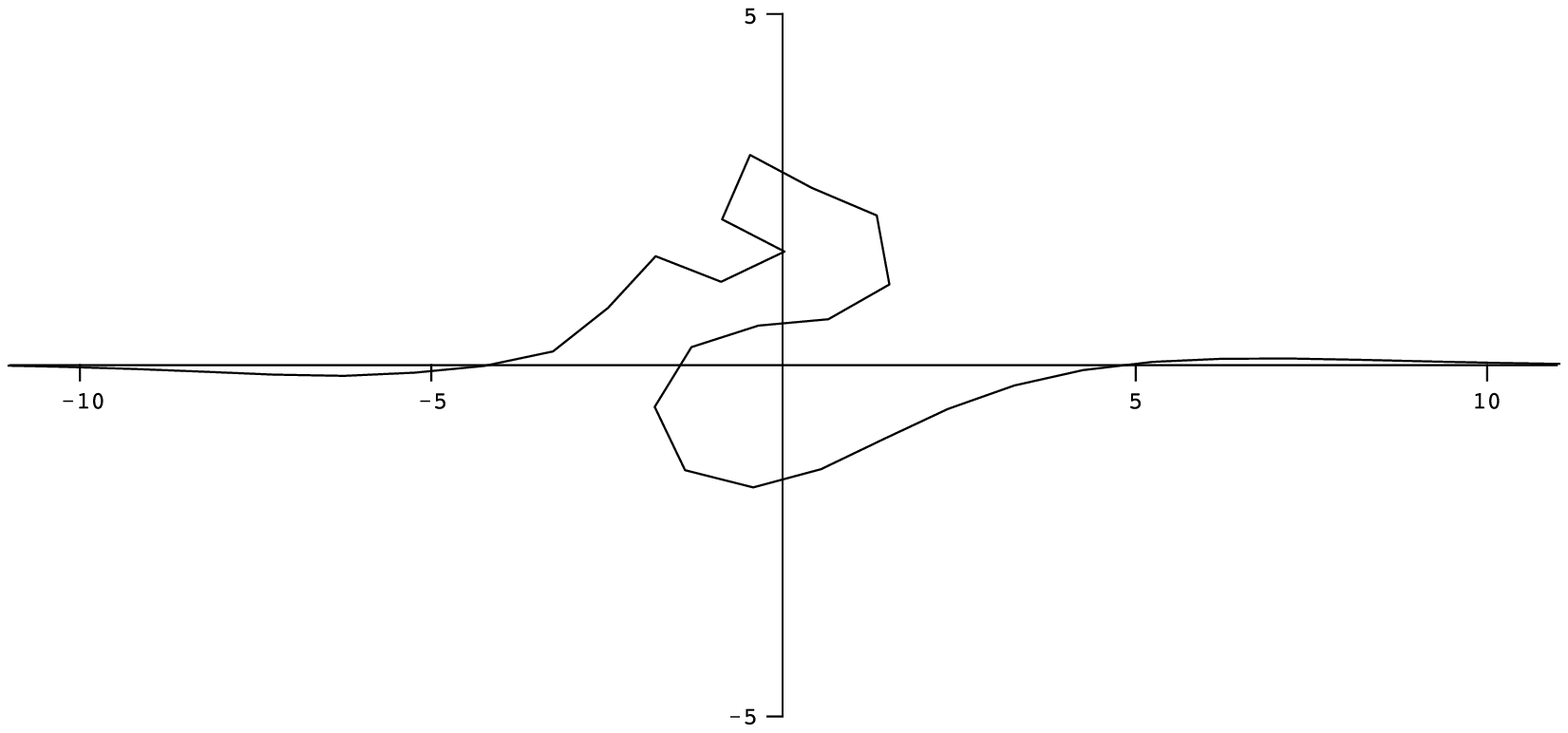}\\
$m = - 7$ 
\end{center}
\end{minipage}
\begin{minipage}{.45\textwidth}
\begin{center}
\includegraphics[width=7truecm]{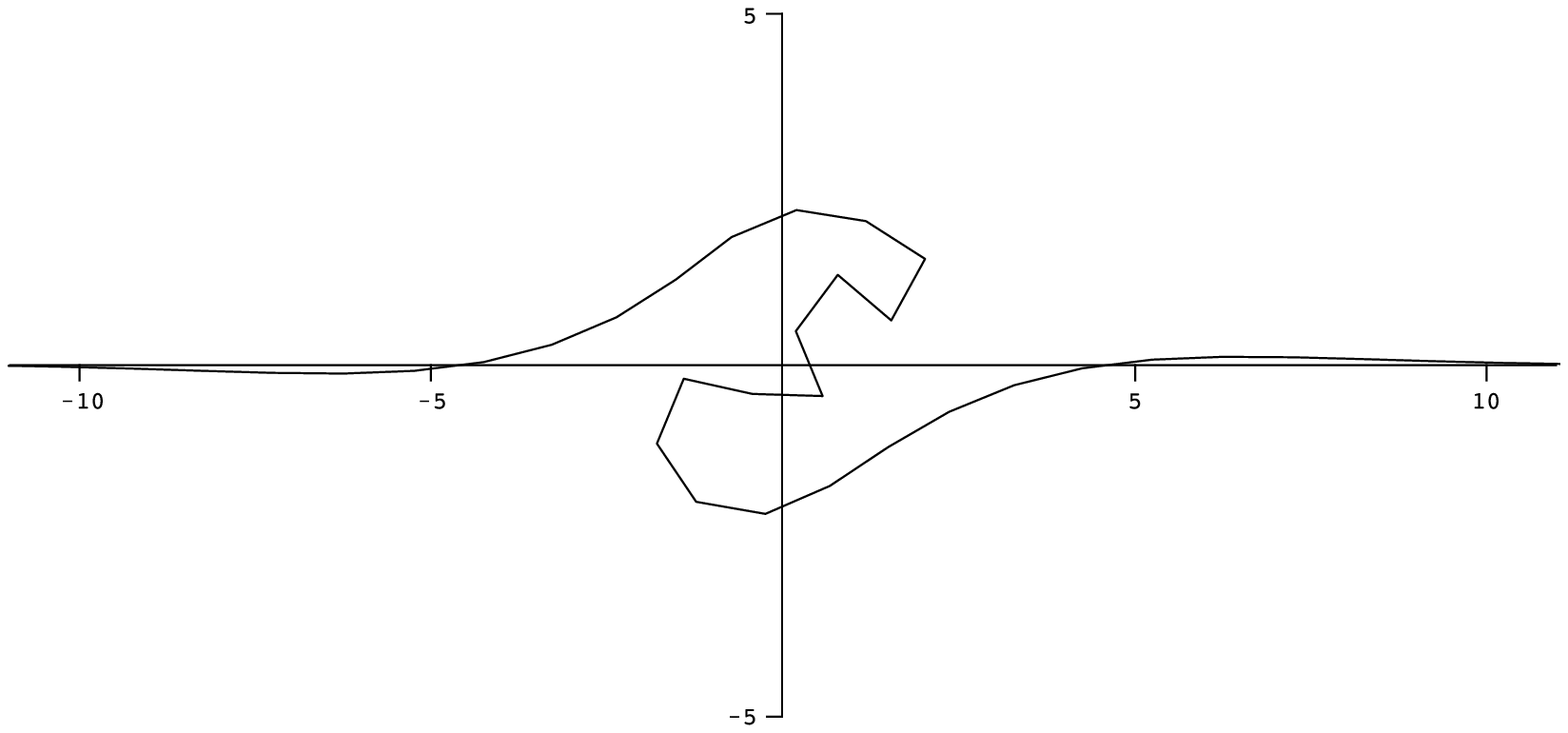}\\
$m = - 1$ 
\end{center}
\end{minipage}\hfill
\begin{minipage}{.45\textwidth}
\begin{center}
\includegraphics[width=7truecm]{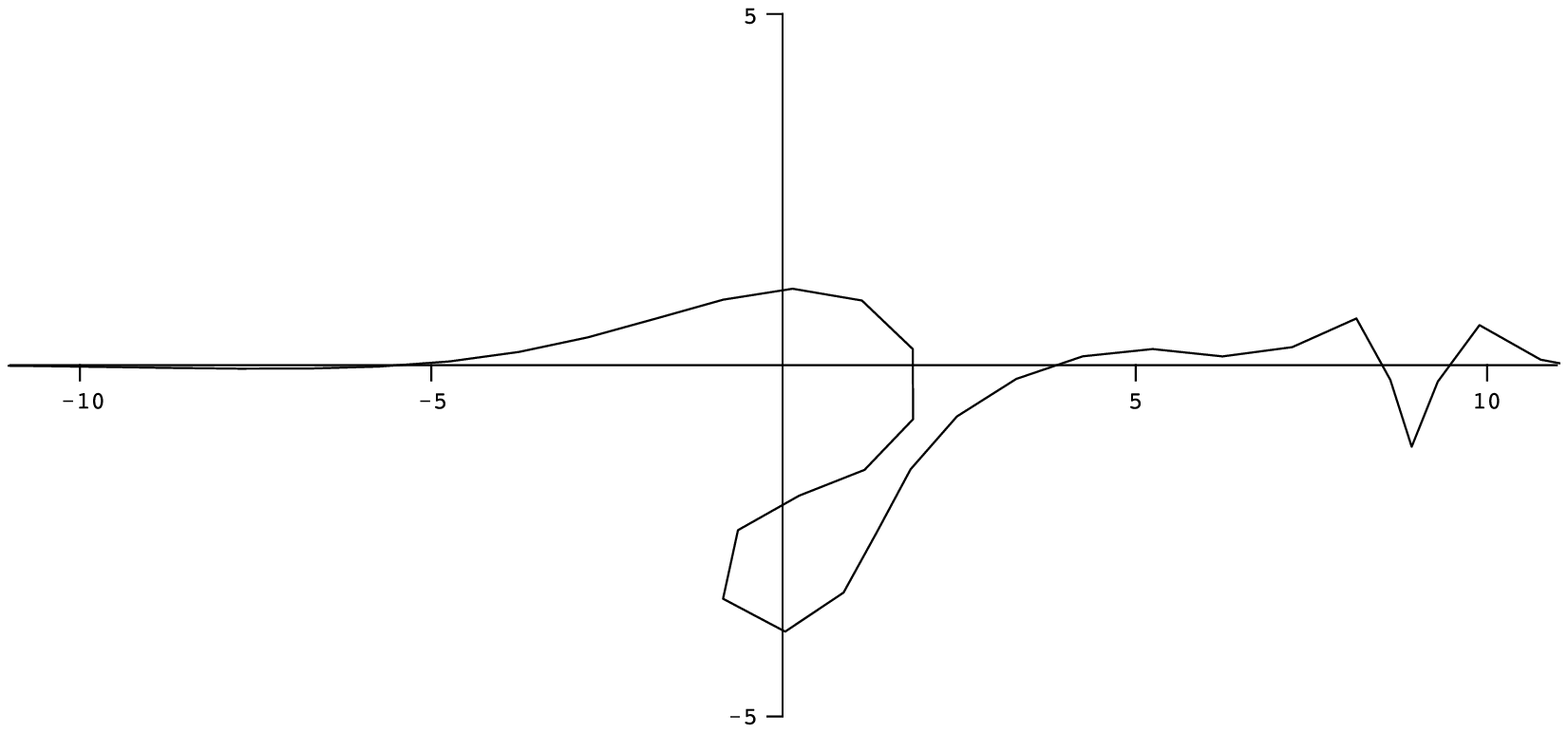}\\
$m = 20$ 
\end{center}
\end{minipage}
\caption{Parameters in equations (\ref{Casorati}), (\ref{tau:Casorati_entries0}) and
(\ref{tau:Casorati_entries}): $N=4$, $x=0$, $y=0$, 
$\alpha_{1}=\alpha_{2}^*=1$, 
$\alpha_{3}=\alpha_{4}^*=\sqrt{-1}$, $\beta_{1}=-\beta_{2}^*=1$,
$\beta_{3}=-\beta_{4}^*=\sqrt{-1}$,
$p_1=p_2^*=0.2-0.2\sqrt{-1}$, $p_3=p_4^*=0.8+0.8\sqrt{-1}$, $a_n=1$,
$b_m=1.5$.}\label{fig:2-breather}
\end{figure}

We present some pictures of the motions of the discrete curves. Figure \ref{fig:1-soliton} shows the
simplest example of a curve, which corresponds to the 1-soliton solution (loop soliton). 
The next example illustrated in Figure \ref{fig:2-soliton} describes the interaction of two loops,
which corresponds to the 2-soliton solution.
Figures \ref{fig:1-breather} and \ref{fig:2-breather} show the motions 
which correspond to the 1-breather and 2-breather solutions, respectively.

\subsection{Solutions via B\"acklund transformations}
In the theory of integrable systems, the B\"acklund transformations are obtained from the shift of a
certain discrete independent variable, which also applies to our geometric transformations.  We
first introduce discrete variables $k$, $l$, and regard the determinant size $N$ as an additional
discrete variable. We then extend the $\tau$ function as $\tau_n^m(k,l,N)=\tau_n^m(x,t;y;k,l,N)$ in
the following way:
\begin{equation}
 \tau_n^m(k,l,N) = \exp\left[-\left(x + \sum_{n'}^{n-1}a_{n'} + \sum_{m'}^{m-1}b_{m'}
+ \sum_{k'}^{k-1}c_{k'}+ \sum_{l'}^{l-1}\frac{1}{d_{l'}}\right)y\right]
~\det\left(f_{j-1}^{(i)}\right)_{i,j=1,\ldots,N},\label{Casorati_BT} 
\end{equation}
\begin{equation}
f_s^{(i)} = e^{\eta_i} + e^{\mu_i},\label{tau:Casorati_entries0_BT}
\end{equation}
\begin{equation}
\left\{
\begin{array}{l}
 {\displaystyle  e^{\eta_i} = \alpha_ip_i^s
  \prod_{n'}^{n-1}(1-a_{n'}p_i)^{-1}\prod_{m'}^{m-1}(1-b_{m'}p_i)^{-1}
\prod_{k'}^{k-1}(1-c_{k'}p_i)^{-1}\prod_{l'}^{l-1}\left(1-\frac{d_{l'}}{p_i}\right)^{-1}}\\
\hskip100pt{\displaystyle \times~e^{p_i x - 4p_i^3 t +  \frac{1}{p_i}y }},\\[4mm]
{\displaystyle  e^{\mu_j}
= \beta_i (-p_i)^s \prod_{n'}^{n-1}(1+a_{n'}p_i)^{-1}\prod_{m'}^{m-1}(1+b_{m'}p_i)^{-1}
\prod_{k'}^{k-1}(1+c_{k'}p_i)^{-1}\prod_{l'}^{l-1}\left(1+\frac{d_{l'}}{p_i}\right)^{-1}}\\
\hskip100pt{\displaystyle \times~e^{-p_ix  + 4p_i^3 t -  \frac{1}{p_i}y }}.
\end{array}\right.
\label{tau:Casorati_entries_BT}
\end{equation}
Accordingly, we extend the relevant dependent variables such as $\Theta$ and $\gamma$ in the same way.
%
\begin{proposition}\label{prop:BT3}\hfill
 \begin{enumerate}
  \item For any $k\in\mathbb{Z}$, $\widetilde{\gamma}(x,t)=\gamma(x,t;k+1)$ is a B\"acklund
	transform of $\gamma(x,t)=\gamma(x,t;k)$ related by equation (\ref{BT_continuous}) with $\lambda=\frac{1}{c_k}$.
  \item For any $k\in\mathbb{Z}$, $\widetilde{\gamma}_n^m=\gamma_n^m(k+1)$ is a B\"acklund
	transform of $\gamma_n^m=\gamma_n^m(k)$ related by equation (\ref{BT_discrete}) with $\lambda=\frac{1}{c_k}$.
  \item For any $N\in\mathbb{Z}_{\geq 0}$, $\widetilde{\gamma}(x,t)=\gamma(x,t;N+1)$ is a B\"acklund
	transform of $\gamma(x,t)=\gamma(x,t;N)$ related by equation (\ref{BT_continuous}) with
	$\lambda=-p_{N+1}$.
  \item For any $N\in\mathbb{Z}_{\geq 0}$, $\widetilde{\gamma}_n^m=\gamma_n^m(N+1)$ is a B\"acklund
	transform of $\gamma_n^m=\gamma_n^m(N)$ related by equation (\ref{BT_discrete}) with $\lambda=-p_{N+1}$.
  \item For any $l\in\mathbb{Z}$, $\overline{\gamma}(x,t)=\gamma(x,t;l+1)$ is a B\"acklund
	transform of $\gamma(x,t)=\gamma(x,t;l)$ related by equation (\ref{BT2:gamma}) with $\lambda=d_{l}$.
  \item For any $l\in\mathbb{Z}$, $\overline{\gamma}_n^m=\gamma_n^m(l+1)$ is a B\"acklund
	transform of $\gamma_n^m=\gamma_n^m(l)$ related by equation (\ref{BT2:gamma_discrete}) with $\lambda=d_{l}$.
 \end{enumerate}
\end{proposition}
\noindent\textbf{Proof.}\quad We first prove (1) and (2). 
It follows from equation (\ref{bl4}) that the $\tau$ function satisfies the
bilinear equation
\begin{equation}
 D_y~\tau_{n}^m(k+1)\cdot\tau_n^m(k) = -c_k\tau^*{}_{n}^m(k+1)\tau^*{}_n^m(k),\label{bl7}
\end{equation}
because of the symmetry with respect to the discrete variables $m$, $n$ and $k$ in
equations (\ref{Casorati_BT})--(\ref{tau:Casorati_entries_BT}). Then by an argument similar to that in
the proof of Theorem \ref{thm:tau_formula}, we see that
\begin{equation}
 \frac{\gamma(k+1)-\gamma(k)}{c_k}
=\left(\begin{array}{l}\medskip
{\cos\left(\frac{\theta(k+1)+\theta(k)}{2}\right) }\\
{\sin\left(\frac{\theta(k+1)+\theta(k)}{2}\right)}\end{array}\right).
\end{equation}
{}From equation (\ref{angle_function}), we have equation (\ref{BT_continuous}) with $\widetilde\gamma=\gamma(k+1)$
and $\widetilde{\theta}=\theta(k+1)$:
\begin{equation}
 \frac{\gamma(k+1)-\gamma(k)}{c_k}=R\left(\frac{\theta(k+1)-\theta(k)}{2}\right)
\gamma'(k).
\end{equation}
Similarly from equation (\ref{segment_n}), we obtain equation (\ref{BT_discrete}) with
$\widetilde\gamma_n^m=\gamma_n^m(k+1)$ and $\widetilde{\Theta}_n^m=\Theta_n^m(k+1)$:
\begin{equation}
 \frac{\gamma_n^m(k+1)-\gamma_n^m(k)}{c_k}=R\left(\frac{\Theta_n^m(k+1)-\Theta_{n+1}^m(k)}{2}\right)
 \frac{\gamma_{n+1}^m(k)-\gamma_n^m(k)}{a_n},
\end{equation}
which proves (1) and (2). The statements (3)--(4) and (5)--(6) can be proved in much the same way
as (1)--(2), by using the bilinear equations
\begin{align}
& D_y~\tau_n^m(N+1)\cdot\tau_n^m(N)=\frac{1}{p_{N+1}}\taus_n^m(N)\taus_n^m(N+1),\label{bl8}\\[2mm]
& D_y~\tau_n^m(l+1)\cdot\taus_n^m(l) = -\frac{1}{d_l}\taus_n^m(l+1)\tau_n^m(l),\label{bl9}
\end{align}
respectively. These bilinear equations will be proved in Appendix. \qquad$\square$
\begin{remark}\rm
Here we give a physical interpretation of the B\"acklund transformations described above.  The
B\"acklund transforms in (1)--(2) and (5)--(6) of Proposition \ref{prop:BT3} correspond to changing
the phase of solitons (loops), in other words, the positions of solitons. On the other hand, the
B\"acklund transforms in (3)--(4) correspond to increasing the number of solitons (loops).
\end{remark}
Computing the potential functions of the B\"acklund transforms of the curves, one can verify the
following result.
\begin{corollary}\hfill
 \begin{enumerate}
  \item For any $k\in\mathbb{Z}$, $\widetilde{\theta}(x,t)=\theta(x,t;k+1)$ is a B\"acklund
	transform of $\theta(x,t)=\theta(x,t;k)$ related by equations (\ref{BT1:pmKdV}) and
	(\ref{BT2:pmKdV}) with $\lambda=\frac{1}{c_k}$.
  \item For any $k\in\mathbb{Z}$, $\widetilde{\Theta}_n^m=\Theta_n^m(k+1)$ is a B\"acklund
	transform of $\Theta_n^m=\Theta_n^m(k)$ related by equations (\ref{BT1:dpmKdV}) and
	(\ref{BT2:dpmKdV}) with $\lambda=\frac{1}{c_k}$.
  \item For any $N\in\mathbb{Z}_{\geq 0}$, $\widetilde{\theta}(x,t)=\theta(x,t;N+1)$ is a B\"acklund
	transform of $\theta(x,t)=\theta(x,t;N)$ related by equations (\ref{BT1:pmKdV}) and
	(\ref{BT2:pmKdV}) with $\lambda=-p_{N+1}$.
  \item For any $N\in\mathbb{Z}_{\geq 0}$, $\widetilde{\Theta}_n^m=\Theta_n^m(N+1)$ is a B\"acklund
	transform of $\Theta_n^m=\Theta_n^m(N)$ related by equations (\ref{BT1:dpmKdV}) and
	(\ref{BT2:dpmKdV}) with $\lambda=-p_{N+1}$.
  \item For any $l\in\mathbb{Z}$, $\overline{\theta}(x,t)=\theta(x,t;l+1)$ is a B\"acklund
	transform of $\theta(x,t)=\theta(x,t;l)$ related by equations (\ref{BT4:pmKdV}) and
	(\ref{BT5:pmKdV}) with $\lambda=d_{l}$.
  \item For any $l\in\mathbb{Z}$, $\overline{\Theta}{}_n^m=\Theta_n^m(l+1)$ is a B\"acklund
	transform of $\Theta_n^m=\Theta_n^m(l)$ related by equations (\ref{BT3:dpmKdV}) and
	(\ref{BT4:dpmKdV}) with $\lambda=d_{l}$.
 \end{enumerate}
\end{corollary}

\appendix
\makeatletter
\@addtoreset{equation}{section}
\renewcommand{\theequation}{\@Alph\c@section.\@arabic\c@equation}
\makeatother
\section{Derivation of bilinear equations (\ref{bl8}) and (\ref{bl9})}
In this appendix, we show that the $\tau$ function given in
equations (\ref{Casorati_BT})--(\ref{tau:Casorati_entries_BT}) actually satisfies the bilinear equations
(\ref{bl8}) and (\ref{bl9}). For this purpose, we first introduce the generic $\tau$ function 
$\tau_n^m(k,l,N;s)=\tau_n^m(x,t;y,z;k,l,N;s)$ by
\begin{equation}
 \tau_n^m(k,l,N;s) 
= \exp\left[-\left(x + \sum_{n'}^{n-1}a_{n'} + \sum_{m'}^{m-1}b_{m'}
+ \sum_{k'}^{k-1}c_{k'}+ \sum_{l'}^{l-1}\frac{1}{d_{l'}}
\right)y\right]
~\det\left(f_{s+j-1}^{(i)}\right)_{i,j=1,\ldots,N},\label{Casorati:without_reduction_BT}
\end{equation}
for $(x,t;y,z)\in\mathbb{R}^4$, $(m,n,k,l,s)\in\mathbb{Z}^5$ and $N\in\mathbb{Z}_{\geq 0}$.
We require $f_{s}^{(i)}=f_{s}^{(i)}(x,t;y,z;m,n;k,l,N)$ ($i=1,\ldots,N$) to satisfy 
the linear equations (\ref{Casorati:linear_continuous}), (\ref{Casorati:linear_discrete})
and 
\begin{equation}
 \frac{f_s^{(i)}(k,l)-f_s^{(i)}(k-1,l)}{c_{k-1}} = f_{s+1}^{(i)}(k,l),\quad
 \frac{f_s^{(i)}(k,l)-f_s^{(i)}(k,l-1)}{d_{l-1}} = f_{s-1}^{(i)}(k,l).\label{Casorati:linear_discrete_BT}
\end{equation}
A typical example for $f_s^{(i)}$ is given by
\begin{equation}
 f_s^{(i)} = e^{\eta_i} + e^{\mu_i},\label{f_without_reduction_BT}
\end{equation}
\begin{equation}
\left\{
\begin{array}{l}
{\displaystyle  e^{\eta_i}=\alpha_i p_i^s
\prod_{n'}^{n-1}(1-a_{n'}p_i)^{-1}\prod_{m'}^{m-1}(1-b_{m'}p_i)^{-1}
\prod_{k'}^{k-1}(1-c_{k'}p_i)^{-1}\prod_{l'}^{l-1}\left(1-\frac{d_{l'}}{p_i}\right)^{-1}
e^{p_i x - 4p_i^3 t +  \frac{1}{p_i}y }},\\[4mm]
{\displaystyle  e^{\mu_j}= \beta_i q_i^s 
\prod_{n'}^{n-1}(1-a_{n'}q_i)^{-1}\prod_{m'}^{m-1}(1-b_{m'}q_i)^{-1}
\prod_{k'}^{k-1}(1-c_{k'}q_i)^{-1}\prod_{l'}^{l-1}\left(1-\frac{d_{l'}}{q_i}\right)^{-1}}\\
\hskip100pt{\displaystyle \times~e^{q_ix  - 4q_i^3 t +  \frac{1}{q_i}y
}},
\end{array}\right.
\label{Casorati_entries:without_reduction_BT}
   \end{equation}
where $p_i$,  $q_i$, $\alpha_i$ and $\beta_i$ are arbitrary complex constants.
We put 
\begin{equation}
 \sigma_n^m(y;k,l,N;s)=\det\left(f_{s+j-1}^{(i)}\right)_{i,j=1,\ldots,N}.
\end{equation}
\begin{proposition}\label{prop:bl_appendix}
The function $\sigma$ satisfies the following bilinear equations:
\begin{align}
& D_y~\sigma_n^m(N+1;s)\cdot\sigma_n^m(N;s) = \sigma_n^m(N;s+1)\sigma_n^m(N+1;s-1),\label{appendix:bl1}\\[2mm]
& \left(D_y-\frac{1}{d_l}\right)~\sigma_n^m(l+1;s)\cdot\sigma_n^m(l;s+1) 
= -\frac{1}{d_l}\sigma_n^m(l+1;s+1)\sigma_n^m(l;s).\label{appendix:bl2}
\end{align}
\end{proposition}
We apply the determinantal technique in order to prove Proposition \ref{prop:bl_appendix}.  The
bilinear equations are reduced to the Pl\"ucker relations, which are quadratic identities of
determinants whose columns are appropriately shifted.  To this end, we construct such formulas that
express the determinants in the Pl\"ucker relations in terms of the derivative or shift of a discrete
variable of $\sigma_n^m(k,l,N;s)$ by using the linear relations of the entries. For the details of
the technique, we refer to \cite{Hirota:book,OKMS:RT,OHTI:dKP,MKO:dRT,MO:dNLS_dark}.

We introduce the notation 
\begin{equation}
\sigma_n^m(l,N;s)
=\left|~ 0_l,\ 1_l,\ \cdots,\ N-2_l,\ N-1_l~\right|,\label{appendix:formula1}
\end{equation}
where `$j_l$' denotes the column vector
\begin{equation}
 j_{{l}} = \left[\begin{array}{c} f_{s+j}^{(1)}(l)
\\\vdots\\f_{s+j}^{(N)}(l) \end{array}\right].
\end{equation}
\begin{lemma}\label{lem:appendix_differential_formula}
 The following formulas hold:
\begin{align}
 &\partial_y \sigma_n^m(l,N;s) 
 =\left|~ -1,\ 1,\ \cdots,\ N-2,\ N-1~\right|,\label{appendix:formula2}\\[2mm]
 &\sigma_n^m(l+1,N;s)=\left|~ 0_{l+1},\ 1,\ \cdots,\ N-2,\ N-1~\right|,\label{appendix:formula3}\\[2mm]
 &d_l\sigma_n^m(l+1,N;s)=\left|~ 1_{l+1},\ 1,\ \cdots,\ N-2,\ N-1~\right|,\label{appendix:formula4}\\[2mm]
 & -\left(d_l\partial_y-1\right) \sigma_n^m(l+1,N;s)
=\left|~ 0,\ 1_{l+1},\ 2,\ \cdots,\ N-2,\ N-1~\right|.\label{appendix:formula5}
\end{align}
Note that the subscripts of column vectors are shown only when $l$ is shifted for notational simplicity.
\end{lemma}
\noindent\textbf{Proof}. Equation  (\ref{appendix:formula2}) can be verified by direct calculation by
using the fourth equation in (\ref{Casorati:linear_continuous}). We have 
\begin{equation}
\sigma_n^m(l+1,N;s)  = \left|~ 0_{l+1},\ 1_{l+1},\ \cdots,\ N-2_{l+1},\ N-1_{l+1}~\right|.
\end{equation}
Adding the $(N-1)$th column multiplied by $d_{l}$ to the $N$th column and 
using equation (\ref{Casorati:linear_discrete_BT}), we have
\begin{equation}
\sigma_n^m(l+1,N;s)  = \left|~ 0_{l+1},\ 1_{l+1},\ \cdots,\ N-2_{l+1},\ N-1_{l}~\right|.
\end{equation}
Similarly, adding the $(i-1)$th column multiplied by $d_{l}$ to the $i$th column and 
using equation (\ref{Casorati:linear_discrete_BT}) for $i=N-1,\ldots,2$, we obtain
\begin{equation}
\sigma_n^m(l+1,N;s)  = \left|~ 0_{l+1},\ 1,\ \cdots,\ N-2,\ N-1~\right|,
\end{equation}
which is equation  (\ref{appendix:formula3}).  Multiplying $d_{l}$ to the first column of
equation (\ref{appendix:formula3}) and using equation (\ref{Casorati:linear_discrete_BT}), we obtain
equation (\ref{appendix:formula4}). Finally, differentiating equation (\ref{appendix:formula4}) with respect to
$y$ yields
\begin{align}
 d_l\partial_y\sigma_n^m(l+1,N;s)&=
\left|~ 0_{l+1},\ 1,\ 2,\ \cdots,\ N-2,\ N-1~\right| 
+ \left|~ 1_{l+1},\ 0,\ 2,\ \cdots,\ N-2,\ N-1~\right|\nonumber\\
&= \sigma_n^m(l+1,N;s) - \left|~ 0,\ 1_{l+1},\ 2,\ \cdots,\ N-2,\ N-1~\right|,
\end{align}
which is equivalent to equation (\ref{appendix:formula5}). This completes the proof. \qquad$\square$
\par\bigskip

\noindent\textbf{Proof of Proposition \ref{prop:bl_appendix}}\quad 
Consider the Pl\"ucker relation (see, for example, \cite{OKMS:RT}),
\begin{equation}
\begin{split}
0=& \left|~-1,0,1,\cdots,N-2~\right|\times\left|~1,\cdots,N-2,N-1,\phi~\right|\\
+& \left|~0,1,\cdots,N-2,N-1~\right|\times\left|~-1,1,\cdots,N-2,\phi~\right|\\
-&  \left|~0,1,\cdots,N-2,\phi~\right|\times\left|~-1,1,\cdots,N-2,N-1~\right|,
\label{appendix:pl1} 
\end{split}
\end{equation}
where $\phi$ is a column vector given by
\begin{equation}
 \phi = \left[\begin{array}{c}0\\\vdots\\0\\1\end{array}\right].
\end{equation}
By using equations (\ref{appendix:formula1}) and (\ref{appendix:formula2}), expanding the determinant with
respect to the column $\phi$, equation (\ref{appendix:pl1}) can be rewritten as
\begin{equation}
0= \sigma_n^m(N;s-1)~\sigma_n^m(N-1;s+1) 
+ \sigma_n^m(N;s)~\partial_y\sigma_n^m(N-1;s)
- \sigma_n^m(N-1;s)~\partial_y\sigma_n^m(N;s),
\end{equation}
which implies equation (\ref{appendix:bl1}). Similarly, applying 
Lemma \ref{lem:appendix_differential_formula}
on the Pl\"ucker relation
\begin{equation}
\begin{split}
0=& \left|~-1, 0,1,\cdots,N-2~\right|\times\left|~0_{l+1},1,\cdots,N-2,N-1~\right|\\
- & \left|~0_{l+1},0,1,\cdots,N-2~\right|\times\left|~-1, 1,\cdots,N-2,N-1~\right|\\
- & \left|~ 0,1,\cdots,N-2,N-1~\right|\times\left|~-1,0_{l+1},1,\cdots,N-2~\right|,
\label{appendix:pl2} 
\end{split}
\end{equation}
we obtain
\begin{equation}
\begin{split}
0&= \sigma_n^m(l;s-1)\times\sigma_n^m(l+1;s) 
-d_l\sigma_n^m(l+1;s-1)\times\partial_y\sigma_n^m(l;s)\\
&-  \sigma_n^m(l;s)\times\left[-(d_l\partial_y-1)\sigma_n^m(l+1;s-1)\right],
\end{split}
\end{equation}
which is equivalent to equation (\ref{appendix:bl2}). This completes the proof. \qquad$\square$
\par\bigskip

{}From Proposition \ref{prop:bl_appendix} and equation (\ref{Casorati:without_reduction_BT}), we see that
$\tau_n^m(k,l,N;s)$ satisfies
\begin{align}
& D_y~\tau_n^m(N+1;s)\cdot\tau_n^m(N;s) = \tau_n^m(N;s+1)\tau_n^m(N+1;s-1),\label{appendix:bl3}\\[2mm]
& D_y~\tau_n^m(l+1;s)\cdot\tau_n^m(l;s+1) 
= -\frac{1}{d_l}\tau_n^m(l+1;s+1)\tau_n^m(l;s).\label{appendix:bl4}
\end{align}
We finally obtain equations (\ref{bl8}) and (\ref{bl9}) from equations (\ref{appendix:bl3}) and
(\ref{appendix:bl4}), respectively, by imposing the reduction condition (\ref{reduction_condition}).

\section*{Acknowledgements} 
One of the authors (K.K.) would like to thank Professor Tim Hoffmann
for giving a series of introductory lectures on discrete differential 
geometry at Kyushu University. This work is partially supported by
JSPS Grant-in-Aid for Scientific Research No.~19340039, 21540067,
21656027 and 22656026.



\begin{thebibliography}{99}
\bibitem{Ablowitz:book} M. J. Ablowitz B. Prinari and A.D. Trubatch, 
Discrete and Continuous Nonlinear Schr\"odinger Systems (Cambridge University Press, Cambridge, 2004).
\bibitem{Bianchi} L. Bianchi, Sulla trasformazione di B\"acklund per le superficie pseudosferiche,
	Rend. Lincei {\bf 5}(1892) 3--12.
\bibitem{BP} A. Bobenko and U. Pinkall, Discrete surface with constant negative Gaussian curvature
	and the Hirota equation, J. Differential Geom. {\bf 43}(1996) 527--611.
\bibitem{Bobenko-Suris} A.I. Bobenko and Y.B. Suris, Discrete Differential Geometry (American
	Mathematical Society, Prividence, RI, 2008).
\bibitem{DJM:discrete1} E. Date, M. Jimbo and T. Miwa, Method for generating discrete
	soliton equations.I, J. Phys. Soc. Jpn. {\bf 51}(1982) 4116--4124.
\bibitem{DJM:discrete2} E. Date, M. Jimbo and T. Miwa, Method for generating discrete
	soliton equations.II, J. Phys. Soc. Jpn. {\bf 51}(1982) 4125--4131.
\bibitem{DJM:discrete3} E. Date, M. Jimbo and T. Miwa, Method for generating discrete
	soliton equations.III, J. Phys. Soc. Jpn. {\bf 52}(1983) 388--393.
\bibitem{DJM:discrete4} E. Date, M. Jimbo and T. Miwa, Method for generating discrete
	soliton equations.IV, J. Phys. Soc. Jpn. {\bf 52}(1983) 761--765.
\bibitem{DJM:discrete5} E. Date, M. Jimbo and T. Miwa, Method for generating discrete
	soliton equations.V, J. Phys. Soc. Jpn. {\bf 52}(1983) 766--771.
\bibitem{Doliwa:dToda} A. Doliwa, Geometric discretization of the Toda system, Phys. Lett. {\bf
	A234}(1997)187--192.
\bibitem{DS1} A. Doliwa and P.M. Santini, Integrable dynamics of a discrete curve and the
	Ablowitz-Ladik hierarchy, J. Math. Phys. {\bf 36} (1995)1259--1273.
\bibitem{DS2} A. Doliwa and P.M. Santini, The integrable dynamic of a discrete curve, 
	Symmetries and Integrability of Difference Equations, D. Levi, L. Vinet and P. Winternitz
	(eds.), (AMS, Providence 1996) 91--102.
\bibitem{DS3} A. Doliwa and P.M. Santini, Geometry of discrete curves and lattices and integrable
	difference equations, Discrete Integrable Geometry and Physics, A. Bobenko and
	R. Seiler (eds.), (Clarendon Press, Oxford, 1999) 139--154.
\bibitem{Goldstein-Petrich} 
R.~E.~Goldstein and D.~M.~Petrich, 
The Korteweg-de Vries hierarchy as dynamics of closed curves in the plane,
Phys. Rev. Lett. {\bf 67} (1991) 3203--3206. 
\bibitem{Hasimoto} H. Hasimoto, A soliton on a vortex filament, J. Fluid. Mech. {\bf 11}(1972)
	477--485.
\bibitem{Hirota:difference1} 
R. Hirota, Nonlinear partial difference equations. I. A difference analogue of
	the Korteweg-de Vries equation, J. Phys. Soc. Jpn. {\bf 43}(1977) 1429--1433.
\bibitem{Hirota:difference2} 
R. Hirota, Nonlinear partial difference equations. II. Discrete-time Toda equation, 
	 J. Phys. Soc. Jpn. {\bf 43}(1977) 2074--2078.
\bibitem{Hirota:dsG} R. Hirota, Nonlinear partial difference equations. III. Discrete sine-Gordon
	equation, J. Phys. Soc. Jpn. {\bf 43}(1977) 2079--2086.
\bibitem{Hirota:difference4} R. Hirota, Nonlinear partial difference equations. IV. B\"acklund
	transformation for the discrete-time Toda equation, J. Phys. Soc. Jpn. {\bf 45}(1978) 321--332.
\bibitem{Hirota:difference5} R. Hirota, Nonlinear partial difference equations. V. 
Nonlinear equations reducible to linear equations, J. Phys. Soc. Jpn. {\bf 46}(1979) 312--319.
\bibitem{Hirota:dpmKdV} R. Hirota, Discretization of the potential modified KdV equation,
	J. Phys. Soc. Jpn. {\bf 67}(1998) 2234--2236.
\bibitem{Hirota:book} R. Hirota, The Direct Method in Soliton Theory,
Cambridge Tracts in Mathematics {\bf 155} (Cambridge University Press, Cambridge, 2004) 
\bibitem{HNW}
M.~Hisakado, K.~Nakayama and M.~Wadati, 
Motion of discrete curves in the plane,
J. Phys. Soc. Jpn.  {\bf 64}  (1995) 2390--2393. 
\bibitem{Hoffmann:dNLS} T.Hoffmann, Discrete Hashimoto surfaces and a doubly discrete smoke-ring
	flow, Discrete Differential Geometry, A.I. Bobenko, P. Schr\"oder, J.M. Sullivan and
	G.M. Ziegler (eds.), Oberwolfach Seminars Vol.39 (Birkh\"auser, Basel, 2008)95--115.
\bibitem{Hoffmann:LN} T. Hoffmann, Discrete Differential Geometry of Curves and Surfaces,
COE lecture Notes Vol. 18 (Kyushu University, Fukuoka, 2009).
\bibitem{HK}
T.~Hoffmann and N.~Kutz, 
Discrete curves in $\mathbb{C}P^1$ and the Toda lattice,
Stud. Appl. Math.  {\bf 113} (2004) 31--55.       
\bibitem{Jimbo-Miwa} M. Jimbo and T. Miwa, Solitons and infinite dimensional Lie algebras,
	Publ. RIMS {\bf 19}(1983) 943-1001.
\bibitem{Lamb} G.~Lamb Jr., 
Solitons and the motion of helical curves,
Phys. Rev. Lett. {\bf 37} (1976) 235--237. 
\bibitem{MKO:dRT} K. Maruno, K. Kajiwara and M. Oikawa, Casorati determinant
	 solution for the discrete-time relativistic Toda lattice equation, Phys. Lett. {\bf
	A241}(1998) 335--343. 
\bibitem{MO:dNLS_dark} 
K. Maruno and Y. Ohta, Casorati determinant form of dark soliton
	 solutions of the discrete nonlinear Schr\"odinger equation, J. Phys. Soc. Jpn. {\bf 75}(2006) 054002.
\bibitem{Matsuura}
N.~Matsuura,
Discrete KdV and discrete modified KdV equations arising from 
motions of discrete planar curves, Int. Math. Res. Not. IMRN Advance Access published May 15, 2011, doi:10.1093/imrn/rnr080.
\bibitem{Miwa} T. Miwa, On Hirota's difference equations, Proc. Japan Acad. Ser. A Math. Sci. {\bf
	58}(1982) 9--12.
\bibitem{Nijhoff} F.W. Nijhoff and H. Capel, The discrete KdV equation, Acta Appl. Math. {\bf
	39}(1995) 133--158.
\bibitem{OHTI:dKP} Y. Ohta, R. Hirota, S. Tsujimoto and T. Imai,
	 Casorati and discrete Gram type determinant representations of
	 solutions to the discrete KP hierarchy, J. Phys. Soc. Jpn. {\bf
	62}(1993) 1872--1886.
\bibitem{OKMS:RT} Y. Ohta, K. Kajiwara, J. Matsukidaira and J. Satsuma,
	 Casorati determinant solution for the relativistic Toda lattice
	 equation, J. Math. Phys. {\bf 34}(1993) 5190--5204.
\bibitem{Pinkall:dNLS} U. Pinkall, B. Springborn, and S. Wei{\ss}mann, A new doubly discrete
	analogue of smoke ring flow and the real time simulation of fluid flow, J. Phys. A:
	Math. Theor. {\bf 40} (2007) 1256312576.
\bibitem{Rogers-Schief} C. Rogers and W.K. Schief, B\"acklund and Darboux Transformations: Geometry
	and Modern Applications in Soliton Theory, Cambridge Texts in Applied Mathematics (Cambridge
	University Press, Cambridge, 2002).
\bibitem{Suris:book} Y. B. Suris, The Problem of Integrable Discretization (Birkh\"auser, Basel, 2003).
\bibitem{Tsujimoto} S. Tsujimoto, On a discrete analogue of the two-dimensional Toda
	 lattice hierarchy, Publ. RIMS {\bf 38}(2002) 113-133.
\bibitem{Ueno-Takasaki} K. Ueno and K. Takasaki, Toda lattice hierarchy, 
Group Representations and Systems of Differential Equations, Adv. Stud. Pure Math. {\bf 4}
	(Kinokuniya, Tokyo, 1982) 1--95.
\bibitem{Wadati} M. Wadati, 
B\"acklund transformation for solutions of the modified Korteweg-de Vries equation,
J. Phys. Soc. Jpn. {\bf 36}(1974) 1498.
\end{thebibliography}
\end{document}